\documentclass[reprint,aip,pop,graphicx,numerical,amsfonts,amsmath,amssymb,twocolumn]{revtex4-1}

\usepackage{bm}
\usepackage[dvips]{graphicx}
\usepackage{lineno}
\usepackage{color}
\usepackage{soul}
\usepackage{booktabs}
\usepackage{multirow}
\sethlcolor{yellow}

\newcommand{\biota}{\iota
                     \hskip-.15ex{\hbox to 0pt{\hss {\leavevmode
                     \hbox{\raise -.60ex \hbox{{\tt \'{}}}}}}}
                     \hskip.37ex{\hbox to 0pt{\hss {\leavevmode
                     \hbox{\raise -.50ex \hbox{{\tt \'{}}}}}}}}

\begin{document}

\preprint{00}

\title{
Eulerian variational formulations and momentum conservation laws 
for kinetic plasma systems
}



\author{H. Sugama}
\affiliation{
National Institute for Fusion Science, 
Toki 509-5292, Japan
}
\affiliation{
Department of Fusion Science, SOKENDAI (The Graduate University for Advanced Studies), 
Toki 509-5292, Japan 
}

\author{M. Nunami}
\affiliation{
National Institute for Fusion Science, 
Toki 509-5292, Japan
}
\affiliation{
Department of Fusion Science, SOKENDAI (The Graduate University for Advanced Studies), 
Toki 509-5292, Japan 
}

\author{S. Satake}
\affiliation{
National Institute for Fusion Science, 
Toki 509-5292, Japan
}
\affiliation{
Department of Fusion Science, SOKENDAI (The Graduate University for Advanced Studies), 
Toki 509-5292, Japan 
}

\author{T.-H. Watanabe}
\affiliation{
Department of Physics,
Nagoya University,  
Nagoya 464-8602, Japan
}


\date{\today}

\begin{abstract}

The Eulerian variational principle for 
the Vlasov-Poisson-Amp\`{e}re system of equations 
in a general coordinate system is presented.
   The invariance of the action integral under an arbitrary spatial 
coordinate transformation is used to 
obtain the momentum conservation law and the symmetric 
pressure in a more direct way than using the translational 
and rotational symmetries of the system. 
Next, the Eulerian variational principle is given 
for the collisionless drift kinetic equation,
where 
particles' phase-space trajectories in given electromagnetic fields 
are described by Littlejohn's guiding center 
equations~[R. G. Littlejohn, J. Plasma Phys.\ {\bf 29}, 111 (1983)]. 
  Then, it is shown that, in comparison with the conventional moment method, 
the invariance under a general spatial coordinate transformation 
yields a more convenient way to obtain 
the momentum balance as a three-dimensional vector equation
in which the symmetric pressure tensor, the Lorentz force, and  
the magnetization current are properly expressed. 
   Furthermore, the Eulerian formulation is presented for 
the extended drift kinetic system, for which, 
in addition to the drift kinetic equations for the distribution functions 
of all particle species, 
the quasineutrality condition and Amp\`{e}re's law 
to determine the self-consistent electromagnetic fields are given.
  Again, the momentum conservation law for the extended system 
is derived from the invariance under the general spatial coordinate transformation. 
  Besides,  the momentum balances are investigated for 
the cases where the collision and/or external source terms are 
added into the Vlasov and drift kinetic equations.     
\end{abstract}


\maketitle 



\section{INTRODUCTION}

    So far, a large number of numerical simulations have been performed to 
investigate neoclassical and turbulent transport in toroidal 
plasmas.~\cite{Lin,Dimits,Garbet} 
    As a modern theoretical technique for deriving basic kinetic model equations of 
such simulations, the variational 
principle~\cite{Littlejohn,goldstein,Sugama2000,B&H} is 
used
because the derived equations possess favorable conservation properties 
for long-time simulations to pursue evolutions of plasma profiles resulting 
from transport processes. 
   Also, useful numerical schemes for plasma simulation satisfying  
the conservation properties have been developed 
by directly utilizing the variational formulation 
rather than numerically approximating 
the basic equations derived from the variational 
principle.~\cite{Qin,Kraus,Bottino,Morrison} 
   In recent years, background flow profiles are regarded as one of key factors which influence magnetic plasma confinement and 
large-scale gyrokinetic simulations are actively done to 
investigate momentum transport processes which determine the flow 
profiles.~\cite{Wang,Sarazin,Abiteboul,Idomura2017} 
   Thus, pressure tensors or momentum transport fluxes 
need to be accurately evaluated because 
they play a critical role for the momentum balance 
in both neoclassical and turbulent transport 
theories.~\cite{
Hinton,Sugama1998,Parra2010,Scott,Sugama2011,Krommes,Abel,Sugama2017,SugamaRMPP} 

In Ref.~\cite{Sugama2000}, the Lagrangian variational formulation 
for the electromagnetic gyrokinetic system is presented from an approximate 
reduction of the Vlasov-Poisson-Amp\`{e}re system which 
is equivalent to the Vlasov-Darwin system~\cite{kaufman}
in which such rapid phenomena as the electromagnetic waves with 
the speed of light $c$  
can be removed from the system (the terminology `Vlasov-Poisson-Amp\`{e}re system' has been customarily used instead of `Vlasov-Darwin system' in the literature on the gyrokinetic theories~\cite{Sugama2000,Hahm1988}). 
   It is shown for the Vlasov-Poisson-Amp\`{e}re system 
that, in the presence of the magnetic field, 
the canonical momentum conservation law derived 
from the space translational symmetry contains the asymmetric pressure tensor. 
In Ref.~\cite{Sugama2013}, the angular momentum conservation law derived from 
the rotational symmetry and 
additional complicated procedures of the Belinfante-Rosenfeld 
type~\cite{Dixon} 
were used to obtain the symmetric pressure tensor from 
the asymmetric canonical pressure tensor and to 
derive the same momentum conservation law as given in 
Ref.~\cite{kaufman}. 

   In this work, 
the variational formulations for the Vlasov-Poisson-Amp\`{e}re system and 
the drift kinetic system are presented in the invariant forms under general spatial coordinate transformations in analogy with the theory of 
general relativity.~\cite{Landau} 
For this purpose, the variational formulations here are 
completely based on the Eulerian 
picture~\cite{Newcomb,Cendra,Marsden,Brizard2000,Squire} 
in which the spatial-coordinate dependence of 
the particle and field parts of the Lagrangian density  
can be more equally treated than in another type of formulation 
using the Lagrangian picture partially for 
the particle part.~\cite{Sugama2000,Sugama2013,Low} 
Detailed descriptions about Lagrangian and Eulerian variational 
formulations 
are found in a recent paper by 
Brizard and Tronci~\cite{B&T}. 
The Eulerian method, which was pioneered by Newcomb~\cite{Newcomb} 
to formulate the magnetohydrodynamics equations and is used in 
the present paper, is also called  
the Euler-Poincar\'{e} reduction 
procedure recently~\cite{Cendra,Marsden,Squire,B&T}.  
Here, in our Eulerian formulation, 
all the governing equations for these systems also 
take the invariant forms and the invariance of the action integrals
can be utilized to derive 
the momentum conservation laws and/or the momentum balances  
 as three-dimensional vector equations. 
  The resultant momentum balance equations contain 
the symmetric pressure tensors which 
have $3\times 3$ symmetric matrix components. 
These symmetric pressure tensor components are derived from taking 
the variation of the Lagrangian density with respect to the metric tensor 
components which appear due to the use of the general spatial coordinate 
system. 
The symmetry of the resultant pressure tensor is a natural result 
because the metric tensor is symmetric. 
Thus, the derivation of the symmetric pressure tensors shown 
in the present paper is more direct than the 
Belinfante-Rosenfeld-type technique and other previous methods. 
   Furthermore, 
for all systems considered here, 
not only the momentum conservation laws but also 
the Belinfante-Rosenfeld type formulas~\cite{Sugama2013,Dixon} 
relating 
the symmetric pressure tensors to the 
asymmetric canonical pressure tensors are simultaneously derived 
from the invariance of the action integrals 
under general spatial coordinate transformations.    
 
   It is also found that the formulation presented 
here for deriving the momentum 
conservation law is more convenient than the conventional 
method based on taking moments of the basic kinetic equation
especially for the drift kinetic system.~\cite{Sugama2016}  
Normally, only the component of the momentum balance equation 
in the direction parallel to the magnetic field is 
derived from the parallel moment of the drift kinetic equation
although it is
not trivial what moment 
should be taken for the gyrophase-averaged distribution 
function to  
obtain the perpendicular momentum balance.
  On the other hand, the method based on the invariance with 
respect to the general spatial coordinate transformation 
can be applied to derive 
the momentum balance equations in both 
parallel and perpendicular directions 
simultaneously even for the drift kinetic system. 

   Normally, based on Noether's theorem,~\cite{goldstein} 
the momentum conservation law in a certain direction 
is derived when a given system has a translational symmetry in that direction. 
  Here, it should be noted that the invariance under the general spatial 
coordinate transformation 
holds more generally than the translational symmetry. 
  Even in the case where the latter property is not satisfied,  
the former property can be valid and used to derive 
the momentum balance equation which does not take 
a conservative form. 
   As shown in Sec.~III,  
  the drift kinetic system in given electromagnetic fields 
corresponds to the above-mentioned case.  
 Thus, the momentum balance equation can be obtained 
for the drift kinetic system with general magnetic geometry. 
When self-consistent electromagnetic fields are treated as 
the solutions of the equations 
given simultaneously with the drift kinetic equations 
from the variational principle, 
the explicit dependence on the spatial coordinates is removed 
from the action integral, and accordingly
the momentum conservation law is derived for the total system 
consisting of the charged particles and fields [see Sec.~IV]. 

The rest of this paper is organized as follows.
In Sec.~II, the Eulerian formulation of the variational principle 
for the Vlasov-Poisson-Amp\`{e}re system is presented.
There, the same results as in Ref.~\cite{Sugama2013} are 
reproduced although the general coordinates are used 
to write the equations in the invariant form and 
derive the momentum conservation law in a more direct way 
than in Ref.~\cite{Sugama2013}. 
   In Sec.~III, the Eulerian variational principle 
is applied to the drift kinetic system, for which
the collisionless drift kinetic equation and 
the momentum balance equation are obtained. 
In this system, which is immersed in the strong magnetic field, 
trajectories of charged particles are described by Littlejohn's 
guiding center equations.~\cite{Littlejohn} 
  In Sec.~IV, the variational principle for the drift kinetic system 
is extended so that the quasineutrality condition 
and Amp\`{e}re's law can be derived simultaneously with 
the drift kinetic equations for all particle species to 
determine the electromagnetic fields self-consistently with 
the distribution functions. 
The momentum conservation law for this extended drift kinetic 
system is derived as well. 
  In Sec.~V, it is shown how the momentum conservation and balance derived in 
Sec.~II--IV are modified when the collision terms are added into the basic kinetic 
equations there. 
   Finally, conclusions are given in Sec.~VI. 
In Appendix~A, 
the Eulerian variational principle is presented for the Vlasov-Poisson system 
and its momentum balance is derived. 
The energy conservation law in the Vlasov-Poisson system 
is also obtained in Appendix~B. 
In Appendix~C,  
the energy balance equation and the energy conservation law are 
shown for the drift kinetic systems described in Secs.~III and IV. 

\section{Vlasov-Poisson-Amp\`{e}re system}

Here, the Vlasov-Poisson-Amp\`{e}re system~\cite{Sugama2013} 
is considered as an example of kinetic systems, for which the Eulerian variational principle is presented.
Also, it is shown for this system how to obtain 
the momentum conservation law from 
the invariance of the action integral 
under general coordinate transformations.

\subsection{Eulerian formulation of the variational principle 
in general coordinates}

The distribution function  
on the phase space for particle species $a$
is denoted by 
$F_a (x^i, v^i, t)$ where $(x^i)_{i=1,2,3}$ and  $(v^i)_{i=1,2,3}$ 
are the position and velocity coordinates of the particle, 
respectively, and 
the number of particles of species $a$ in 
the phase-space volume element 
$d^3 x d^3 v \equiv dx^1 dx^2 dx^3 dv^1 dv^2 dv^3$ 
is given by $F_a (x^i, v^i, t) d^3 x d^3 v$. 
Here, $(x^i)_{i=1,2,3}$ represent a general spatial coordinate system 
which can be either a Cartesian or any other curved coordinate system. 
However, in the present paper, we assume that 
the position vector ${\bf r}$ is a function of only the spatial coordinates 
$(x^i)_{i=1,2,3}$ and it is independent of time $t$. 
In the given spatial coordinate system, 
$(v^i)_{i=1,2,3}$ are defined as contravariant components of 
the velocity vector by using 
$(\partial {\bf r}/\partial x^i )_{i=1,2,3}$ as the basis vectors. 

     In the Lagrangian picture, 
the motion of a particle of species $a$ in 
the phase space is described by 
representing the position and velocity of the particle at time $t$ 
as the functions, 
\begin{equation}
\label{Lmotion}
[ x_{a L}^i (x_0^n, v_0^n, t_0 ; t),  v_{a L}^i (x_0^n, v_0^n, t_0 ; t) ]
, 
\end{equation}
which satisfy
the initial conditions at time $t_0$, 
\begin{equation}
 x_{a L}^i (x_0^n, v_0^n, t_0 ; t_0 ) = x_0^i 
, 
\hspace*{3mm}
 v_{a L}^i (x_0^n, v_0^n, t_0 ; t_0 ) = v_0^i 
. 
\end{equation}
Using the Lagrangian representations of the particle's motion 
given in Eq.~(\ref{Lmotion}), 
the distribution function at time $t$ is related to 
that at time $t_0$ by 
\begin{eqnarray}
\label{Faxv}
F_a (x^i, v^i, t) 
& = & 
\int d^3 x_0 \int d^3 v_0
\; 
F_a (x_0^m, v_0^m, t_0) 
\nonumber \\ & & 
\mbox{}
\times 
\delta^3 [x^i - x_{aL}^i (x_0^m, v_0^m, t_0; t) ]
\nonumber \\ & & 
\mbox{}
\times 
\delta^3 [v^i - v_{aL}^i (x_0^m, v_0^m, t_0; t) ]
. 
\end{eqnarray}

   We next represent
the particle's velocity and 
acceleration in the Eulerian picture by  
\begin{equation}
\label{Eulerxv}
u_{ax}^i (x^m,  v^m,  t)
,
\hspace*{3mm}
u_{av}^i (x^m,  v^m,  t)
,
\end{equation}
which are related to those in the Lagrangian picture by 
\begin{eqnarray}
\label{uxvi} 
& & 
\hspace*{-5mm}
u_{ax}^i (
x_{a L}^m (x_0^n, v_0^n, t_0 ; t), 
v_{a L}^m (x_0^n, v_0^n, t_0 ; t), 
t)
\nonumber \\ 
& & =
\dot{x}_{a L}^i (x_0^n, v_0^n, t_0 ; t)
,
\nonumber \\ 
& & 
\hspace*{-5mm}
u_{av}^i (
x_{a L}^m (x_0^n, v_0^n, t_0 ; t), 
v_{a L}^m (x_0^n, v_0^n, t_0 ; t), 
t)
\nonumber \\ 
& & =
\dot{v}_{a L}^i (x_0^n, v_0^n, t_0 ; t)
. 
\end{eqnarray}
     Here, $\dot{f} = \partial f (x_0^m, v_0^m,  t)/ \partial t$ 
stands for the time derivative of an arbitrary function 
$f (x_0^m, v_0^m, t)$ with $(x_0^m, v_0^m)$ kept fixed. 
    Using Eqs.~(\ref{Faxv}) and (\ref{uxvi}),  we can show that the distribution function 
$F_a (x^i, v^i, t)$ satisfies 
the continuity equation in the six-dimensional phase space, 
\begin{equation}
\label{Vlasov0}
\frac{\partial F_a}{\partial t} 
+ \frac{\partial}{\partial x^i}
( F_a u_{a x}^i  )
+ \frac{\partial}{\partial v^i}
( F_a u_{a v}^i )
= 0
. 
\end{equation}
    In the present paper, 
we use the summation convention that 
an index repeated in a term [such as seen in Eq.~(\ref{Vlasov0})]  
represents summation over the range $\{ 1, 2, 3 \}$.

The action integral $I$ to describe the 
Vlasov-Poisson-Amp\`{e}re system is written as
\begin{equation}
\label{I}
I  \equiv  \int_{t_1}^{t_2} dt \; L 
\equiv \int_{t_1}^{t_2}  dt \int_V d^3 x\; {\cal L} 
,
\end{equation}
where the Lagrangian $L$ is defined by the spatial integral of 
the Lagrangian density ${\cal L}$ over the volume $V$  
and  
${\cal L}$ is given by 
\begin{eqnarray}
\label{Ldens} 
{\cal L} 
& \equiv & 
\sum_a 
\int d^3 v \; F_a (x^i, v^i, t) L_a  +  {\cal L}_f  
.
\end{eqnarray}
    Here, the single-particle Lagrangian $L_a$ for species $a$ 
is written in the Eulerian picture as 
\begin{eqnarray}
& & 
\hspace{-8mm}
\label{spL}
L_a [v^i, u_{ax}^i (x^n, v^n, t),  \phi (x^n, t), A_i (x^n, t), g_{ij} (x^n) ]
\nonumber \\ 
& = & 
\left[ m_a g_{ij} (x^n) v^i + \frac{e_a}{c} A_j (x^n, t) \right]  u_{ax}^j (x^n, v^n, t)
\nonumber \\  & & \mbox{}
- \left[ \frac{1}{2} m_a g_{ij} (x^n) v^i v^j  + e_a \phi (x^n, t) \right]
. 
\end{eqnarray}
Note that the single-particle Lagrangian given by Eq.~(2) in 
Ref.~\cite{Sugama2013} 
is reproduced from Eq.~(\ref{spL}) when 
replacing $x^i$, $v^i$, and $u_{ax}^i$ in Eq.~(\ref{spL}) with 
the corresponding Lagrangian representations $x_{aL}^i$, $v_{aL}^i$, 
and $\dot{x}_{aL}$. 

   The field Lagrangian density ${\cal L}_f$ on the right-hand 
side of Eq.~(\ref{Ldens}) is given by 
\begin{eqnarray}
\label{Lf}
& & 
\hspace{-5mm}
{\cal L}_f 
\left[\frac{\partial \phi (x^n, t)}{\partial x^i},  A_i (x^n, t), 
\frac{\partial A_j (x^n, t)}{\partial x^i}, 
\lambda (x^n, t), g_{ij} (x^n), 
\frac{\partial g_{jk} (x^n) }{\partial x^i}
\right]
\nonumber \\ 
&  & 
=
\sqrt{g(x^n)}
\left[ 
\frac{g^{ij}(x^n)}{8\pi}
\frac{\partial \phi (x^n, t)}{\partial x^i}
\frac{\partial \phi (x^n, t)}{\partial x^j}
- \frac{g_{ij}(x^n)}{8\pi}  
\right. 
\nonumber \\ 
&  & 
\left. \mbox{}
\times  B^i (x^n, t) B^j (x^n, t)
+ \frac{\lambda (x^n, t)}{4\pi c}  g^{ij}(x^n) 
\nabla_i A_j (x^n, t) 
\right]
.
\end{eqnarray}
Equation~(\ref{Lf}) is obtained by using the general 
phase-space coordinates  $(x^i, v^i)$ to express 
the field Lagrangian density given by Eq.~(3) 
in Ref.~\cite{Sugama2013}. 

   The contravariant components $(B^i)_{i=1,2,3}$  
of the magnetic field 
are expressed in terms of the covariant components 
$(A_i)_{i=1,2,3}$  of the vector potential as 
\begin{equation}
\label{Bi}
B^i (x^n, t) = \frac{\epsilon^{ijk}}{\sqrt{g(x^n) }} 
\frac{\partial A_k (x^n, t)}{\partial x^j}
\end{equation}
and the components $\nabla_i A_j$ $(i,j=1,2,3)$ of the covariant derivative of 
the covariant vector $A_j$ are defined by
\begin{equation}
\nabla_i A_j (x^n, t)  = 
\frac{\partial A_j (x^n, t)}{\partial x^i}
 - \Gamma_{ij}^k (x^n) A_k (x^n, t)
, 
\end{equation}
where the Levi-Civita symbol is denoted by  
\begin{eqnarray}
\label{eijk}
& & 
\epsilon^{ijk} 
 \equiv   \epsilon_{ijk}
\nonumber \\
&  & 
\equiv
\left\{
\begin{array}{cl}
1 
&
\mbox{($(i, j, k) = (1,2,3), (2,3,1), (3,1,2)$)} 
\\
-1
& 
\mbox{($(i, j, k) = (1,3,2), (2,1,3), (3,2,1)$)} 
\\
0 
& 
\mbox{(otherwise)}, 
\end{array}
\right. 
\end{eqnarray}
the determinant of the metric tensor matrix 
is given by  
\begin{equation}
g (x^n) \equiv  \det [ g_{ij} (x^n) ]
, 
\end{equation}
and 
the Christoffel symbols $\Gamma_{ij}^k$ $(i,j,k=1,2,3)$ are defined 
by~\cite{Schutz}
\begin{eqnarray}
& & \Gamma_{ij}^k (x^n)    \equiv  
 g^{kl} (x^n) \Gamma_{l, ij}(x^n)  
\nonumber \\ 
& & \equiv 
\frac{1}{2} g^{kl}(x^n) 
\left[ 
\frac{\partial g_{jl} (x^n) }{\partial x^i}
+ \frac{\partial g_{li} (x^n) }{\partial x^j}
- \frac{\partial g_{ij} (x^n) }{\partial x^l}
\right]
. 
\hspace*{5mm}
\end{eqnarray}
The covariant and contravariant components of the metric tensor 
components are denoted by $g_{ij}$ and $g^{ij}$, respectively, 
and they satisfy 
\begin{equation}
g^{ik} g_{kj} = \delta^i_j
, 
\end{equation}
where $\delta^i_j$ represents the Kronecker delta defined by 
\begin{equation}
\delta^i_j
\equiv 
\left\{
\begin{array}{cc}
1 & (i = j) \\
0 & (i \neq j) .
\end{array}
\right. 
\end{equation}

We now consider the virtual displacement 
of the particle's trajectory in the phase space, 
which is represented by the variations of the Lagrangian 
representations of the particle's position and velocity 
 in Eq.~(\ref{Lmotion}) 
as 
\begin{equation}
\label{deltaxvL}
\delta  x_{a L} ^i (x_0^m, v_0^m, t_0 ; t ) 
, 
\hspace*{3mm}
 \delta v_{a L}^i (x_0^m, v_0^m, t_0 ; t ) 
. 
\end{equation}
   The variations in the position and velocity are 
represented in the Eulerian picture by 
\begin{equation}
\label{deltaxvE}
\delta  x_{a E}^i (x^m, v^m, t) 
, 
\hspace*{3mm}
\delta  v_{a E}^i (x^m, v^m, t) 
, 
\end{equation}
which are related to those in the Lagrangian picture 
by 
\begin{eqnarray}
\label{deltaxvLE}
& & 
\hspace*{-5mm}
\delta  x_{a E}^i (
x_{a L}^m (x_0^n, v_0^n, t_0 ; t), 
v_{a L}^m (x_0^n, v_0^n, t_0 ; t), 
t)
\nonumber \\ 
& & =
\delta  x_{a L}^i (x_0^n, v_0^n, t_0 ; t)
,
\nonumber \\ 
& & 
\hspace*{-5mm}
\delta v_{a E}^i (
x_{a L}^m (x_0^n, v_0^n, t_0 ; t), 
v_{a L}^m (x_0^n, v_0^n, t_0 ; t), 
t)
\nonumber \\ 
& & =
\delta  v_{a L}^i (x_0^n, v_0^n, t_0 ; t)
. 
\end{eqnarray}
Making use of Eq.~(\ref{uxvi}) to consider the variations in the particle's velocity 
and acceleration which result from the virtual displacement denoted by 
Eq.~(\ref{deltaxvL}), 
we obtain 
\begin{eqnarray}
\label{duvax}
\delta u_{ax}^i
& = & 
\left( \frac{\partial }{\partial t}
+ u_{ax}^j \frac{\partial }{\partial x^j}
+ u_{av}^j \frac{\partial }{\partial v^j}
\right) \delta x_{a E}^i
\nonumber \\ 
& & \mbox{}
- \left( \delta x_{a E}^j \frac{\partial }{\partial x^j}
+ \delta v_{a E}^j \frac{\partial }{\partial v^j}
\right) u_{ax}^i
, 
\nonumber \\ 
\delta u_{av}^i
& = & 
\left( \frac{\partial }{\partial t}
+ u_{ax}^j \frac{\partial }{\partial x^j}
+ u_{av}^j \frac{\partial }{\partial v^j}
\right) \delta v_{a E}^i
\nonumber \\ 
& & \mbox{}
- \left( \delta x_{a E}^j \frac{\partial }{\partial x^j}
+ \delta v_{a E}^j \frac{\partial }{\partial v^j}
\right) u_{av}^i
,
\end{eqnarray}
where Eq.~(\ref{deltaxvLE}) is used as well. 
Here, 
$\delta u_{ax}^i$ and $\delta u_{av}^i$ represent the variations 
in the functional forms of $u_{ax}^i$ and $u_{av}^i$, respectively, 
and the parts of variations in $u_{ax}^i$ and $u_{av}^i$ caused by 
the changes in their arguments are not included in 
$\delta u_{ax}^i$ and $\delta u_{av}^i$. 
   W also find from Eqs.~(\ref{Faxv}) and (\ref{deltaxvLE}) that 
the variation in the distribution function due to 
the virtual displacement of the trajectory shown in 
Eq.~(\ref{deltaxvL}) is given by 
\begin{equation}
\label{dFa}
\delta F_a
=
- \frac{\partial}{\partial x^j}
( F_a \delta x_{a E}^j  )
- \frac{\partial}{\partial v^j}
( F_a \delta v_{a E}^j  )
. 
\end{equation}

We further consider that the spatial functional forms of 
the electrostatic potential $\phi$, 
the covariant components $A_i$ of the vector potential, and 
the field $\lambda$ associated with the Coulomb gauge condition
[see Eq.~(\ref{Coulomb})] are 
virtually varied by $\delta \phi$, $\delta A_i$, and $\delta \lambda$ 
in addition to the virtual displacement of the 
particle's phase-space trajectory. 
   Consequently, the action integral defined by Eq.~(\ref{I}) 
with Eq.~(\ref{Ldens}) is varied by 
\begin{eqnarray}
\label{deltaI}
\delta I
& = & 
\sum_a
\int_{t_1}^{t_2} dt 
\int d^3 x \int d^3 v \; F_a 
\left[
\delta x_{aE}^i \left\{ 
\left( \frac{\partial L_a}{\partial x^i} \right)_{u_{ax}}
\right. \right. 
\nonumber \\
& & \left. \left. \mbox{}
-  \left(  \frac{d}{d t } \right)_a
\left( \frac{\partial L_a}{\partial u_{ax}^i} \right)
\right\}
+ \delta v_{aE}^i 
\left(\frac{\partial L_a}{\partial v^i}\right)_{u_{ax}}
\right]
\nonumber \\
& & \mbox{}
\hspace*{-4mm}
+ \int_{t_1}^{t_2} dt \int d^3 x 
\left[
\delta \phi \left(
- \sum_a e_a \int d^3 v \; F_a 
- \frac{\sqrt{g}}{4\pi} \Delta \phi 
\right)
\right. 
\nonumber \\
& & 
 \mbox{}
+ \delta A_i \left(
\sum_a \frac{e_a}{c} \int d^3 v \; F_a  u_{ax}^i
- \frac{\epsilon^{ijk}}{4\pi}  \frac{\partial B_k}{\partial x^j} 
\right.
\nonumber \\
& & 
\left. 
\left. \mbox{}
- \frac{\sqrt{g} }{4\pi c} g^{ij} \frac{\partial \lambda}{\partial x^j} 
\right)
+ \delta \lambda \frac{\sqrt{g} }{4\pi c}  \nabla_i A^i
\right]
+ \delta I_b
, 
\end{eqnarray}
where  
$( \partial L_a/\partial x^i )_{u_{ax}}$ and 
$( \partial L_a/\partial v^i )_{u_{ax}}$ represent 
the derivatives of $L_a$ in $x^i$ and $v^i$, respectively, 
with $u_{ax}^i$ kept fixed in $L_a$. 
The definitions of the operators $(d/dt)_a$ and $\Delta$ are
shown later in Eqs.~(\ref{ddt}) and (\ref{poisson}), respectively, 
and 
\begin{eqnarray}
\delta I_b
& = & 
\sum_a
\int_{t_1}^{t_2} dt 
\int d^3 x \int d^3 v  
\left[
\frac{\partial}{\partial t } \left(
F_a \frac{\partial L_a}{\partial u_{ax}^i }  \delta x_{aE}^i
\right)
\right.
\nonumber \\
& &
\hspace*{-10mm}
\left. \mbox{}
+ \frac{\partial}{\partial x^j} \left(
F_a u_{ax}^j \frac{\partial L_a}{\partial u_{ax}^i }  \delta x_{aE}^i
\right)
+ \frac{\partial}{\partial v^j} \left(
F_a u_{av}^j \frac{\partial L_a}{\partial u_{ax}^i }  \delta x_{aE}^i
\right)
\right]
\nonumber \\
& &
\hspace*{-10mm}
 \mbox{}
+ \int_{t_1}^{t_2} dt 
\int d^3 x \frac{\partial}{\partial x^i} 
\left[
\frac{\sqrt{g}}{4\pi} 
\left\{
g^{ij} \frac{\partial \phi}{\partial x^j} \delta \phi
+ 
\left( \frac{\epsilon^{ijk}}{\sqrt{g}} B_j 
+ \frac{\lambda}{c} g^{ik} 
\right) 
\right.
\right.
\nonumber \\ 
& &
\hspace*{-10mm}
 \left. \left. \mbox{}
\times \delta A_k
+ \delta x_E^i 
\left( 
\frac{g^{ij}}{2} 
\frac{\partial \phi}{\partial x^i} 
\frac{\partial \phi}{\partial x^j} 
- \frac{1}{2} B^j B_j
+ \frac{\lambda}{c} \nabla_j A^j
\right)
\right\} \right]
\end{eqnarray}
is the part which can be determined from the values of 
the variations $\delta x_{aE}^i$, $\delta \phi$, and 
$\delta A_i$ on the boundaries of the integral region
because of the divergence theorem.  

We now show that the Vlasov-Poisson-Amp\`{e}re system 
obeys the Eulerian variation principle.
Namely, 
$F_a$, $\phi$, and $A_i$ are determined from the condition that 
$\delta I = 0$ for arbitrary variations $\delta x_{aE}^i$, $\delta v_{aE}^i$, 
$\delta \phi$, $\delta A_i$, and $\delta \lambda$
which vanish on the boundaries of the integral region. 
First, it is found from Eq.~(\ref{deltaI}) that 
$\delta I  / \delta v_{aE}^i = 0$ gives
\begin{equation}
\label{dIdv}
F_a 
\left( \frac{\partial L_a}{\partial v^i}  \right)_{u_{ax}}
=
F_a  m_a g_{ij} ( u_{ax}^j  - v^j ) 
= 0
,
\end{equation}
which is rewritten as 
\begin{equation}
\label{Fuv}
F_a u_{ax}^i = F_a  v^i
. 
\end{equation}
Here, we should note that 
$u_{ax}^i = v^i$ is derived from Eq.~(\ref{Fuv}) 
under the condition that  $F_a \neq 0$. 
However, since $u_{ax}^i$ enters Eq.~(\ref{Vlasov0}) 
in the form of the product $F_a u_{ax}^i$, 
it doesn't cause any trouble to simply write 
\begin{equation}
\label{u=v}
u_{ax}^i =  v^i
, 
\end{equation}
from now on 
instead of Eq.~(\ref{Fuv}) 
even without assuming $F_a \neq 0$. 
This simplification of omitting $F_a$ will also 
be done below in the processes where the equation for 
$u_{av}^i$ [see Eq.~(\ref{Newtoneq})] is derived. 
 
We next use
$
\delta I / \delta x_{aE}^i = 0
$
to obtain
\begin{equation}
\label{dpdt}
\left(\frac{d}{dt}\right)_a p_{ai}
=
\left( \frac{\partial L_a}{\partial x^i} \right)_{u_{ax}}
, 
\end{equation}
where the covariant vector component $p_{ai}$ of the 
canonical momentum and the time derivative 
$(d/dt)_a$ along the motion of the particle of species $a$ 
in the phase space are defined by 
\begin{equation}
p_{ai}
 \equiv  
\left( \frac{\partial L_a}{\partial u_{ax}^i} \right)
\equiv
m_a g_{ij} v^j + \frac{e_a}{c} A_i
, 
\end{equation}
and 
\begin{equation}
\label{ddt}
\left(\frac{d}{dt}\right)_a 
 \equiv  
\frac{\partial}{\partial t } 
+  u_{ax}^k \frac{\partial}{\partial x^k} 
+ u_{av}^k \frac{\partial}{\partial v^k} 
, 
\end{equation}
respectively. 
   From Eq.~(\ref{ddt}), we also have   
\begin{equation}
\left(\frac{d}{dt}\right)_a  x^i
=
 u_{ax}^i 
, \hspace*{5mm}
\left(\frac{d}{dt}\right)_a  v^i
= 
 u_{av}^i 
. 
\end{equation}
  Equation~(\ref{dpdt}) can be rewritten as 
the covariant form of Newton's motion equation 
in the general coordinate system,  
\begin{equation}
\label{muvi}
m_a \left(
u_{av i} + \Gamma_{i, jk} v^j v^k
\right)
= 
e_a \left(
E_i 
+ \frac{1}{c}\sqrt{g} \epsilon_{ijk}
v^j B^k
\right)
, 
\end{equation}
   where the covariant component $E_i$ of the electric field 
is defined by 
\begin{equation}
E_i 
\equiv 
- \frac{\partial \phi}{\partial x^i}
- \frac{1}{c} \frac{\partial A_i}{\partial t}
. 
\end{equation}
   The contravariant form of Newton's motion equation 
is obtained from Eq.~(\ref{muvi}) as  
\begin{equation}
\label{Newtoneq}
m_a \left(
u_{av}^i + \Gamma^i_{jk} v^j v^k
\right)
= 
e_a \left(
E^i 
+ \frac{1}{c}\frac{\epsilon^{ijk}}{\sqrt{g}}
v_j B_k
\right)
. 
\end{equation}
It should be noted that the Levi-Civita symbol 
$\epsilon^{ijk} \equiv \epsilon_{ijk}$ 
[see Eq.~(\ref{eijk})] can be regarded as 
either a contravariant tensor density of weight 1 or a 
covariant tensor density of weight $-1$. 
Then, 
$\sqrt{g} \epsilon_{ijk}$ and $\epsilon^{ijk} / \sqrt{g}$ 
represent covariant and contravariant tensors, respectively, 
which are used in the Lorentz force terms 
on the right-hand side of Eqs.~(\ref{muvi}) and (\ref{Newtoneq}).

Substituting Eqs.~(\ref{u=v}) and (\ref{Newtoneq}) 
into Eq.~(\ref{Vlasov0}) yields 
the Vlasov kinetic equation, 
\begin{eqnarray}
\label{Vlasov}
& & 
\frac{\partial F_a}{\partial t} 
+ \frac{\partial}{\partial x^j}
( F_a v^j  )
\nonumber \\
& & 
\mbox{}
+ \frac{\partial}{\partial v^j}
\left[ F_a 
\left\{ 
- \Gamma^i_{jk} v^j v^k 
+
\frac{e_a}{m_a} \left(
E^i 
+ \frac{1}{c} 
\frac{\epsilon^{ijk}}{\sqrt{g}}
v_j B_k
\right)
\right\}
\right]
= 0
. 
\nonumber \\
& & 
\end{eqnarray}
   As noted after Eq.~(\ref{u=v}), 
$F_a$  appears as a factor in the equations 
$\delta I / \delta x_{aE}^i = \delta I / \delta v_{aE}^i = 0$ 
although it is omitted in writing Eqs.~(\ref{u=v}), (\ref{dpdt}), (\ref{muvi}) 
and (\ref{Newtoneq}). 
This omission of $F_a$ does not make a difference in deriving the Vlasov equation 
in Eq.~(\ref{Vlasov}) by substituting the motion 
equations, Eqs.~(\ref{u=v}) and (\ref{Newtoneq}), 
into Eq.~(\ref{Vlasov0}) 
because $u_{ax}^i$ and $u_{av}^i$ enter 
Eq.~(\ref{Vlasov0}) in the forms of the products 
$F_a u_{ax}^i$ and $F_a u_{av}^i$. 

We use
$
\delta I / \delta \lambda = 0
$
to obtain the Coulomb (or transverse) gauge condition, 
\begin{equation}
\label{Coulomb}
\nabla_i A^i 
\equiv 
\frac{1}{ \sqrt{g} }
 \frac{\partial  
( \sqrt{g}  A^i )}{\partial x^i} 
= 
0
.
\end{equation}
 Poisson's equation is derived from 
$
\delta I / \delta \phi = 0
$
as
\begin{equation}
\label{poisson}
\sqrt{g} \Delta \phi \equiv
 \frac{\partial}{\partial x^i}
\left( \sqrt{g} g^{ij} \frac{\partial \phi}{\partial x^j} 
\right)
= 
-
4\pi
 \sum_a e_a \int d^3 v \; F_a 
, 
\end{equation}
and 
$
\delta I / \delta A_i = 0
$
gives
\begin{equation}
\label{ampere0}
\frac{\epsilon^{ijk}}{ \sqrt{g} }
 \frac{\partial B_k}{\partial x^j} 
+ \frac{g^{ij}}{c}  \frac{\partial \lambda}{\partial x^j} 
= 
\frac{4\pi}{c}
j^i
, 
\end{equation}
where $j^i$ represents the $i$th contravariant 
component of the current density vector 
defined by  
\begin{equation}
j^i
= 
\frac{1}{\sqrt{g} } 
\sum_a e_a \int d^3 v \; F_a  u_{ax}^i
. 
\end{equation}
The transverse (or solenoidal) part of Eq.~(\ref{ampere0}) is 
written  as Amp\`{e}re's law, 
\begin{equation}
\label{ampere}
\frac{\epsilon^{ijk}}{ \sqrt{g} }
 \frac{\partial B_k}{\partial x^j} 
= 
\frac{4\pi}{c}
j_T^i
, 
\end{equation}
where $j_T^i$ represents the $i$th contravariant component of 
the transverse part of the the current density vector. 
Note that an arbitrary vector field ${\bf a}$ can be written as
${\bf a} = {\bf a}_L + {\bf a}_T$, where the 
longitudinal 􏰱(or irrotational􏰑) part ${\bf a}_L$ and 
the transverse 􏰱(or solenoidal􏰑) part ${\bf a}_T$ satisfy
$\nabla \times {\bf a}_L = 0$ and $\nabla \cdot {\bf a}_T = 0$, 
respectively.~\cite{jackson} 

Equations~(\ref{Vlasov}), (\ref{poisson}), and 
(\ref{ampere}) are the governing equations for  
the Vlasov-Poisson-Amp\`{e}re system. 
Thus, the same system of equations as shown in Ref.~\cite{Sugama2013} are
reproduced in the present work although the equations here 
are represented using the general spatial coordinates $(x^i)_{i=1,2,3}$ and 
the contravariant velocity vector components $(v^i)_{i=1,2,3}$. 
Using Eq.~(\ref{poisson}), the longitudinal part of Eq.~(\ref{ampere0}), 
and the charge conservation law obtained from 
Eq.~(\ref{Vlasov}), 
we obtain 
\begin{equation}
\label{jL}
\frac{\partial \lambda}{\partial x^i} = 4\pi j_{Li} =  
- \frac{\partial E_{Li}}{\partial t},
\end{equation}
where $E_{Li} = - \partial  \phi / \partial x^i$ represents 
the longitudinal electric field given by the electrostatic potential. 
Then, we can put~\cite{Sugama2013}
\begin{equation}
\lambda = \frac{\partial \phi}{\partial t}
, 
\end{equation}
which is used hereafter. 
  Then,  we find that 
Eqs.~(\ref{poisson}), (\ref{ampere0}), (\ref{ampere}), and (\ref{jL}) 
give the Darwin model~\cite{kaufman} 
as noted in Refs.~\cite{Sugama2000,Sugama2013}.

\subsection{Transformation of spatial coordinates}

We now consider the transformation of 
the spatial coordinates written as  
\begin{equation}
\label{xprime}
 x'^i 
=
x^i + \xi^i (x^n)
,
\end{equation}
where the infinitesimal variation in the spatial coordinate $x^i$ 
is denoted by $\xi^i (x^n)$
which is regarded as an arbitrary function of only the spatial coordinates. 
Under the transformation of the spatial coordinates, 
the velocity components  $(v^i)_{i=1,2,3}$ are transformed as 
the contravariant vector components. 
Thus, the velocity component  $v'^i$ in the transformed coordinate system 
is written as 
\begin{equation}
 v'^i 
=
\frac{\partial x'^i (x^n)}{\partial x^j} v^j
=
v^i + \overline{\delta} v^i 
, 
\end{equation}
where the infinitesimal variation $\overline{\delta} v^i$ 
in the velocity component is given by 
\begin{equation}
\label{ovldv}
\overline{\delta} v^i 
=
\frac{\partial \xi^i (x^n)}{\partial x^j} v^j
.
\end{equation}
Here and hereafter, 
we use $\overline{\delta} \cdots$ 
 to represent the variation associated with 
the infinitesimal spatial coordinate transformation 
which should be distinguished from 
the variation $\delta  \cdots$ due to the virtual 
displacement treated in Sec.~II.A. 

The electrostatic potential is a scalar which is invariant under  
the transformation of the spatial coordinates, 
\begin{equation}
\label{phi}
\phi' (x'^n, t) 
=
\phi(x^n, t) 
. 
\end{equation}
Here,  we define  the variation $\overline{\delta} \phi$ 
in the functional form of 
$\phi$ due to the spatial coordinate transformation by 
\begin{equation}
\label{dbphi}
\overline{\delta} \phi (x^n, t) 
\equiv 
\phi' (x^n, t) - 
\phi (x^n, t).  
\end{equation}
Note that the spatial arguments of $\phi'$ and $\phi$ are the same as 
each other on the right-hand side of Eq.~(\ref{dbphi}) 
while they are different in Eq.~(\ref{phi}). 
Then, substituting  
$\phi' (x'^n, t) \simeq \phi' (x^n, t) 
+ \xi^i (x^n) \partial  \phi' (x^n, t) /\partial x^i
\simeq \phi' (x^n, t) 
+ \xi^i (x^n) \partial  \phi (x^n, t) /\partial x^i$ 
into Eq.~(\ref{phi}) and 
using Eq.~(\ref{dbphi}), 
we obtain 
\begin{equation}
\label{deltaphi}
 \overline{\delta} \phi (x^n,  t)
=
-  \xi^i (x^n)\frac{\partial \phi (x^n, t)}{\partial x^i} 
\equiv 
- ( L_\xi  \phi ) (x^n,  t)
,
\end{equation}
where $L_\xi$ denotes the Lie derivative~\cite{Marsden2} 
associated with the vector field 
$(\xi^i)$. 
In the same way as in Eq.~(\ref{deltaphi}), 
the variation $\overline{\delta} \lambda$ in another scalar variable $\lambda$ 
is written as 
\begin{equation}
\label{deltalambda}
\overline{\delta}\lambda (x^n,  t)
=
-  \xi^i (x^n)\frac{\partial \lambda (x^n, t)}{\partial x^i} 
\equiv 
- ( L_\xi  \lambda) (x^n,  t)
.
\end{equation}

In the transformed spatial coordinates, 
the covariant vector components of the vector potential are 
written as 
\begin{equation}
\label{Ai}
A'_i (x'^n, t) 
=
\frac{\partial x^j }{\partial x'^i}
A_j (x^n, t) 
. 
\end{equation}
  In the same way as in Eq.~(\ref{dbphi}), 
we define the variation $\overline{\delta} A_i$  
in the functional form of 
$A_i$ due to the spatial coordinate transformation by 
\begin{equation}
\label{dbA}
\overline{\delta} A_i (x'^n, , t) \equiv 
A'_i (x^n, t)  -  A_i (x^n, t)
.
\end{equation}
Substituting the formulas 
$A'_i (x'^n, t) \simeq  A'_i (x^n, t)  + \xi^j (x^n) \partial A_i (x^n, t)/\partial x^j$ 
and 
$\partial x^j /\partial x'^i \simeq 
\delta^j_i  - \partial \xi^j (x^n) /\partial x^i$ 
into Eq.~(\ref{Ai}) 
and using Eq.~(\ref{dbA}), 
we obtain  
\begin{eqnarray}
\label{deltaA}
\overline{\delta} A_i (x^n,  t)
& = & 
 - \xi^j (x^n) \frac{\partial A_i (x^n, t)}{\partial x^j} 
- \frac{\partial \xi^j (x^n)}{\partial x^i}
A_j (x^n, t) 
\nonumber \\ 
& \equiv & 
- ( L_\xi  A_i ) (x^n,  t)
, 
\end{eqnarray}
where we see that 
the Lie derivative $L_\xi$  
can be used again to represent $\overline{\delta} A_i$.  

The contravariant vector components $E^i$, 
the covariant metric tensor components $g_{ij}$, and 
the contravariant tensor components $g^{ij}$ are
transformed as 
\begin{eqnarray}
\label{Egg}
E'^i (x'^n, t) 
& = & 
\frac{\partial x'^i }{\partial x^j}
E^j (x^n, t) 
, 
\nonumber \\
g'_{ij} (x'^n, t) 
& = & 
\frac{\partial x^k }{\partial x'^i}
\frac{\partial x^l }{\partial x'^j}
g_{kl} (x^n, t) 
, 
\nonumber \\
g'^{ij} (x'^n, t) 
& = & 
\frac{\partial x'^i }{\partial x^k}
\frac{\partial x'^j }{\partial x^l}
g^{kl} (x^n, t) 
.  
\end{eqnarray}
Then, following the procedures similar to those used in deriving 
Eqs.~(\ref{deltaphi}) and (\ref{deltaA}), 
the variations in the functional forms of $E^i$, $g_{ij}$, and $g^{ij}$ 
due to the spatial coordinate transformation are 
derived as
\begin{eqnarray}
\label{dEgg}
& & \overline{\delta} E^i 
=
- L_\xi E^i
= 
- \xi^j \frac{\partial E^i }{\partial x^j}
 + \frac{\partial \xi^i }{\partial x^j}
E^j 
, 
\nonumber \\
& & \overline{\delta} g_{ij}
=
- L_\xi g_{ij}
= 
- \xi^k \frac{\partial g_{ij} }{\partial x^k}
- \frac{\partial \xi^k}{\partial x^i} g_{kj}
- \frac{\partial \xi^k}{\partial x^j} g_{ik}
\nonumber \\
& & 
\hspace*{7mm}
=
 - \nabla_i \xi_j - \nabla_j \xi_i
, 
\nonumber \\
& & \overline{\delta} g^{ij}
=
- L_\xi g^{ij}
= 
- \xi^k \frac{\partial g^{ij} }{\partial x^k}
+ \frac{\partial \xi^i}{\partial x^k} g^{kj}
+ \frac{\partial \xi^j}{\partial x^k} g^{ik}
\nonumber \\
& & 
\hspace*{7mm}
= 
\nabla^i \xi^j + \nabla^j \xi^i
. 
\end{eqnarray}

The transformation of the spatial coordinates given by 
Eq.~(\ref{xprime}) changes 
the Lagrangian representations of 
the trajectory of the particle's motion in the phase space as 
\begin{eqnarray}
\label{xvprime}
x'^i_{aL} (x'^n_0, v'^n_0, t_0 ; t)
& = & 
x^i_{aL} (x_0^n, v_0^n, t_0 ; t)
+ \xi^i ( x^m_{aL}(x_0^n, v_0^n, t_0 ; t) )
,
\nonumber \\ 
v'^i_{aL}  (x'^n_0, v'^n_0, t_0 ; t)
& = & 
v^i_{aL} (x_0^n, v_0^n, t_0 ; t)
\nonumber \\  & & \hspace{-8mm} 
\mbox{}
+ \eta^i ( x^m_{aL}(x_0^n, v_0^n, t_0 ; t), v^m_{aL} (x_0^n, v_0^n, t_0 ; t))
, 
\end{eqnarray}
where the particle's position and velocity at time $t_0$ 
are written in the transformed coordinates as
\begin{eqnarray}
\label{xvprime0}
x'^i_0 
& = & 
x^i_0 + \xi^i ( x'^n_0  ) 
,
\nonumber \\ 
v'^i_0 
& = & 
v^i_0 + \eta^i  ( x'^n_0,   v'^n_0 ) 
. 
\end{eqnarray}
Since $v^i$ is the contravariant vector component,  
its variation $\eta^i$ caused by the change $\xi^i$ 
in the spatial coordinate $x^i$ can be written as 
\begin{equation}
 \eta^i  ( x^n,   v^n ) 
= 
\frac{\partial \xi^i (x^n)}{\partial x^j} v^j
. 
\end{equation}

In the transformed coordinate system, 
the distribution function is given by
\begin{eqnarray}
\label{Fprime}
F'_a (x'^i, v'^i, t) 
& = & 
\int d^3 x'_0 \int d^3 v'_0
\; 
F'_a (x'^n_0, v'^n_0, t_0) 
\nonumber \\ & & 
\mbox{}
\times 
\delta^3 [x'^i - x'^i_{aL} (x'^n_0, v'^n_0, t_0; t) ]
\nonumber \\ & & 
\mbox{}
\times 
\delta^3 [v'^i - v'^i_{aL} (x'^n_0, v'^n_0, t_0; t) ]
. 
\end{eqnarray}
   Here, the initial distribution functions $F'_a (x'^n_0, v'^n_0, t_0)$ 
and $F_a (x^n_0, v^n_0, t_0)$ in the transformed and original 
coordinate systems are related to each other by 
$F'_a (x'^n_0, v'^n_0, t_0) d^3 x'_0 d^3 v'_0
= F_a (x^n_0, v^n_0, t_0) d^3 x_0 d^3 v_0$,  
from which 
we obtain 
\begin{equation}
\label{F00}
F'_a (x'^n_0, v'^n_0, t_0) 
= F_a (x^n_0, v^n_0, t_0) 
\left[
\det \left( \frac{\partial x'^i_0}{\partial x^j_0} \right)
\det \left( \frac{\partial v'^i_0}{\partial v^j_0} \right)
\right]^{-1}
. 
\end{equation}
The variation $\overline{\delta} F_a$ in the functional form of 
the distribution function 
due to the spatial coordinate transformation
is defined by 
\begin{equation}
\label{deltaF0}
F'_a (x^n, v^n, t) 
= F_a (x^n, v^n, t) 
+ \overline{\delta} F_a (x^n, v^n, t)
. 
\end{equation}
Then, using Eqs.~(\ref{xvprime}), (\ref{Fprime}), (\ref{F00}), 
and (\ref{deltaF0}), 
we obtain 
\begin{equation}
\label{dFxieta}
\overline{\delta} F_a 
=
- \frac{\partial}{\partial x^j}
( F_a \xi^j  )
- \frac{\partial}{\partial v^j}
( F_a \eta^j  )
. 
\end{equation}

The relations between the Eulerian and Lagrangian representations 
of the particle's velocity and acceleration shown in Eq.~(\ref{uxvi}) 
are rewritten in the transformed coordinate system as
\begin{eqnarray}
\label{uxvprime}
& & u'^i_{ax} ( x'^m_{aL}  ( x'^n_0,   v'^n_0, t_0 ; t),  
v'^m_{aL}  ( x'^n_0,   v'^n_0, t_0 ; t),  t)
\nonumber \\ 
& & =
\dot{x}'^i_{aL} ( x'^n_0,   v'^n_0, t_0 ; t)
,
\nonumber \\ 
& & u'^i_{av} ( x'^m_{aL}  ( x'^n_0,   v'^n_0, t_0 ; t),  
v'^m_{aL}  ( x'^n_0,   v'^n_0, t_0 ; t),  t)
\nonumber \\ 
& & = 
\dot{v}'^i_{aL} ( x'^n_0,   v'^n_0, t_0 ; t)
.  
\end{eqnarray}
We also write 
\begin{eqnarray}
\label{deltauxv}
u'^i_{ax} ( x^n, v^n,  t) 
& =  &
u^i_{ax} ( x^n, v^n,  t)  + \overline{\delta} u^i_{ax} ( x^n, v^n,  t) 
\nonumber \\ 
u'^i_{av} ( x^n, v^n,  t) 
& =  &
u^i_{av} ( x^n, v^n,  t)  + \overline{\delta} u^i_{av} ( x^n, v^n,  t) 
, 
\end{eqnarray}
to define $\overline{\delta} u^i_{ax}$ 
and  $\overline{\delta} u^i_{av}$ as the variations 
in the Eulerian functional forms  
of the particle's velocity and acceleration, respectively. 
   Using Eqs.~(\ref{xvprime}), (\ref{uxvprime}), and (\ref{deltauxv}), 
we find that $\overline{\delta} u^i_{ax}$ 
and $\overline{\delta} u^i_{av}$ are written as 
\begin{eqnarray}
\label{duxieta}
& & 
\overline{\delta} u_{ax}^i
=
u_{ax}^j \frac{\partial \xi^i}{\partial x^j} 
- \left( \xi^j \frac{\partial }{\partial x^j}
+\eta^j \frac{\partial }{\partial v^j}
\right) u_{ax}^i
, 
\nonumber \\ 
& & 
\overline{\delta} u_{av}^i
=
\left( 
u_{ax}^j \frac{\partial }{\partial x^j}
+ u_{av}^j \frac{\partial }{\partial v^j}
\right) \eta^i
- \left( \xi^j \frac{\partial }{\partial x^j}
+\eta^j \frac{\partial }{\partial v^j}
\right) u_{av}^i
. 
\nonumber \\ 
& & 
\end{eqnarray}

\subsection{Derivation of the momentum conservation law}

We can use $F'_a (x'^n, v'^n, t)$, $u_{ax}'^i (x'^n, v'^n, t)$, 
$\phi' (x'^n, t)$, $A'_i (x'^n, t)$, and $g'_{ij} (x'^n, t)$ in 
Eqs.~(\ref{I})--(\ref{Lf})
to define the action integral $I'$ in the transformed coordinates 
$(x'^n, v'^n)$. 
Then, using Eqs.~(\ref{deltaphi}), (\ref{deltalambda}), (\ref{deltaA}), 
(\ref{dEgg}), (\ref{dFxieta}) and (\ref{duxieta}), 
we find that the variation $\overline{\delta} I \equiv I' - I$ in the action integral is 
written as  
\begin{eqnarray}
\label{dI}
\overline{\delta} I
& = & 
\int_{t_1}^{t_2} dt 
\int_V d^3 x 
\left[ 
\xi_j 
\left( 
\frac{\partial P_c^j}{\partial t}
+ \nabla_i \Theta^{ij}
\right) 
\right.
\nonumber \\ 
& & 
\left. \mbox{}
+ \frac{\partial }{\partial x^i}
\left\{ \xi_j  \left(
\Pi_c^{ij} - \Theta^{ij}
-\nabla_k F^{ijk} \right)
\right\} \right]
,
\end{eqnarray}
where the canonical momentum vector density $P_c^j$ 
and the canonical pressure tensor density $\Pi_c^{ij}$ 
are defined by
\begin{eqnarray}
\label{Pcj}
 P_c^j 
& \equiv & 
g^{jk}
\sum_a \int d^3 v \, F_a 
\frac{\partial L_a}{\partial u_{ax}^k}
\nonumber \\
& = & 
\sum_a \int d^3 v \, F_a 
\left( m_a v^j + \frac{e_a}{c} A^j \right)
,
\end{eqnarray}
and 
\begin{eqnarray}
\Pi_c^{ij}
&  \equiv &
g^{jk} (\Pi_c)^i_k
\nonumber \\
 &  \equiv  &
g^{jk}
\left(
\sum_a \int d^3 v \, F_a u_{ax}^i
\frac{\partial L_a}{\partial u_{ax}^k}
+ {\cal L}_f \delta^i_k
\right. 
\nonumber \\
& & 
\left. \mbox{}
-  \frac{\partial {\cal L}_f}{\partial (\partial \phi/\partial x^i)}
\frac{\partial \phi}{\partial x^k}
-  \frac{\partial {\cal L}_f}{\partial (\partial A_l/\partial x^i)}
\nabla_k A_l
\right)
\nonumber \\
&  = &  
\sum_a \int d^3 v \, F_a v^i
\left( m_a v^j + \frac{e_a}{c} A^j \right)
\nonumber \\
& & 
\mbox{}
+ \frac{\sqrt{g}}{8\pi} g^{ij}
\left( E_L^k E_{Lk} - B^k B_k \right)
+ \frac{\sqrt{g}}{4\pi} \left(
- E_L^i E_L^j
\right.
\nonumber \\
& & 
\left. \mbox{}
+ \frac{\epsilon^{ikl}}{\sqrt{g}} B_l \nabla^j A_k
- \frac{1}{c} \frac{\partial \phi}{\partial t} \nabla^j A^i
\right)
,
\end{eqnarray}
respectively. 
   The symmetric tensor density $\Theta^{ij}$ and 
the third-rank tensor density $F^{ijk}$ are defined by
\begin{eqnarray}
\label{thetaij}
\Theta^{ij}
& \equiv & 
2
\left[  
\sum_a \int d^3 v \, F_a 
\frac{\partial L_a}{\partial g_{ij}}
+
\frac{\partial {\cal L}_f}{\partial g_{ij}}
- 
\frac{\partial }{\partial x^k}
\left( 
\frac{\partial {\cal L}_f }{\partial (\partial g_{ij}/ \partial  x^k) }
\right) \right]
\nonumber \\
& = & 
\sum_a \int d^3 v \, F_a m_a v^i v^j + \sqrt{g}
\left[
\frac{g^{ij}}{8\pi} (E_L^k E_{Lk} + B^k B_k) 
\right.
\nonumber \\
&  & 
\mbox{} 
- \frac{1}{4\pi} (E_L^i E_L^j + B^i B^j) 
- \frac{1}{4\pi c} \left(
A^i \frac{\partial E_L^j}{\partial t} 
+
A^j \frac{\partial E_L^i}{\partial t} 
\right.
\nonumber \\
&  & 
\left. \left.
\mbox{} 
- g^{ij} A^k \frac{\partial E_{Lk}}{\partial t} 
\right)
\right]
, 
\end{eqnarray}
and 
\begin{eqnarray}
\label{Fijk}
& & F^{ijk}
 \equiv 
-  A^j 
\frac{\partial {\cal L}_f }{\partial (\partial A_k / \partial  x^i) }
- 2 
\frac{\partial {\cal L}_f }{\partial (\partial g_{jk}/ \partial  x^i) }
\nonumber \\ 
&  & =
\frac{\sqrt{g}}{4\pi} 
\left[
A^j
\left(
\nabla^i A^k - \nabla^k A^i
\right) 
+
\frac{1}{c} \frac{\partial \phi}{\partial t} 
\left(
g^{ij} A^k - g^{jk} A^i
\right)
\right]
, 
\hspace*{8mm}
\end{eqnarray}
respectively. 
  In deriving Eq.~(\ref{dI}), 
Eqs.~(\ref{u=v}), (\ref{dpdt}), (\ref{Coulomb}),  (\ref{poisson}), 
and (\ref{ampere0}), 
which are derived from the variational principle in Sec.~II.A, 
are also used. 

The symmetry condition, 
\begin{equation}
\Theta^{ij}
= \Theta^{ji}
,
\end{equation}
is naturally confirmed in Eq.~(\ref{thetaij}) 
because the symmetric metric tensor density components $g_{ij}$ are 
used for differentiating the Lagrangian density ${\cal L}$ in 
the definition of $\Theta^{ij}$. 
It should be noted that, in Eq.~(\ref{thetaij}),  
partial derivatives with respect to $g_{ij}$ 
need to be carefully done because 
$3\times 3$ metric tensor components $g_{ij}$
are not completely independent of each other 
due to the constraint $g_{ij} = g_{ji}$. 
Here, for an arbitrary function $f$ of $g_{ij}$, 
the notation  $\partial f/ \partial g_{ij}$ 
is defined such that the infinitesimal variations $\delta g_{ij}$ 
in $g_{ij}$ give rise to the variation 
$\delta f = (\partial f/ \partial g_{ij}) \delta g_{ij}$ in $f$
where  both $\delta g_{ij}$ and  $\partial f/ \partial g_{ij}$ 
must be symmetric under exchange of the 
indices $i$ and $j$,~\cite{Landau} 
For example,  
we have  
$\partial g_{kl}/ \partial g_{ij} = \frac{1}{2}
( \delta^i_k \delta^j_l + \delta^j_k \delta^i_l )$
according to the above-mentioned definition. 
In the same manner, 
derivatives with respective to $\partial g_{ij}/\partial x^k$ shown 
in Eqs.~(\ref{thetaij}) and (\ref{Fijk}) are defined taking into account 
the symmetry under exchange of the indices $i$ and $j$. 

We see from Eq.~(\ref{Fijk}) that the third-rank tensor density components 
$F^{ijk}$ is anti-symmetric with respect to exchanging the superscripts $i$ and $k$, 
\begin{equation}
\label{antis}
F^{ijk}
= - F^{kji}
. 
\end{equation}
Using Eq.~(\ref{antis}) and 
the commutation condition, 
\begin{equation}
\label{comm}
\nabla_i \nabla_k
= \nabla_k \nabla_i
, 
\end{equation}
we obtain
\begin{eqnarray}
\label{nablaik}
\nabla_i \nabla_k F^{ijk}
= 0
. 
\end{eqnarray}
Note that the commutation condition in Eq.~(\ref{comm}) 
is valid because the three-dimensional real space considered here 
is a flat one with no curvature. 
For a general curved space,  
the Riemann curvature tensor $R^l_{mik}$ is used to write~\cite{Schutz}  
\begin{equation}
\nabla_i \nabla_k V^l
- \nabla_k \nabla_i V^l
= R^l_{mik} V^m
, 
\end{equation}
where $V^m$ is the $m$th contravariant component of an arbitrary vector field.  
  Then, we find 
\begin{eqnarray}
\label{ddF}
& & 
\nabla_i \nabla_k F^{ijk}
- \nabla_k \nabla_i F^{ijk}
\nonumber \\
& = & 
R^i_{mik} F^{mjk} + R^j_{mik} F^{imk} + R^k_{mik} F^{ijm}
\nonumber \\
& = & 
- R_{mk} F^{mjk}  + R_{mi} F^{ijm}
= 0
, 
\end{eqnarray}
where 
Eq.~(\ref{antis}) and the properties of the Riemann tensor 
($R^a_{bcd} = - R^a_{bdc}$,  $R_{bc} \equiv R^a_{bca} = R_{cb}$) 
are used. 
  Thus, we find the interesting fact that 
Eq.~(\ref{nablaik}) is valid even in the curved space when 
Eq.~(\ref{antis}) is satisfied. 

Since the action integral is invariant under 
an arbitrary transformation of the spatial coordinates, 
$\overline{\delta} I$ shown in Eq.~(\ref{dI}) vanishes for any $\xi_j$ 
so that we obtain 
the momentum conservation law, 
\begin{equation}
\label{momcons}
\frac{\partial P_c^j}{\partial t}
+ \nabla_i \Theta^{ij}
= 0 
, 
\end{equation}
and the relation of the symmetric pressure tensor density 
$\Theta^{ij}$ to the asymmetric canonical tensor density 
$\Pi_c^{ij}$, 
\begin{equation}
\label{thetapi}
\Theta^{ij}
= 
\Pi_c^{ij} 
-\nabla_k F^{ijk} 
. 
\end{equation}
   Equations~(\ref{momcons}) and (\ref{thetapi}) are
derived from the conditions that the integrands at the interior and boundary   
points shown on the right-hand side of Eq.~(\ref{dI}) should vanish, respectively. 
   Combining Eq.~(\ref{nablaik}) with Eq.~(\ref{thetapi}) leads to 
\begin{equation}
\label{nablaT0}
\nabla_i \Theta^{ij}
= 
\nabla_i \Pi_c^{ij}
, 
\end{equation}
 which can be used to rewrite the momentum conservation law 
in Eq.~(\ref{momcons}) as 
\begin{equation}
\label{momcons2}
\frac{\partial P_c^j}{\partial t}
+ \nabla_i \Pi_c^{ij}
= 0 
. 
\end{equation}
In Ref.~\cite{Sugama2013}, 
the momentum conservation law, Eq.~(\ref{momcons2}), including 
the asymmetric canonical momentum tensor density, $\Pi_c^{ij}$, 
is derived from the space translational symmetry of the action integral $I$ 
before the relation of the Belinfante-Rosenfeld type symmetric pressure tensor 
$\Theta^{ij}$ to $\Pi_c^{ij}$ in Eq.~(\ref{thetapi}) is obtained 
from the rotational symmetry of $I$ to derive 
the other momentum conservation law, Eq.~(\ref{momcons}).  
   On the other hand, in the present work, 
both the momentum conservation law, Eq.~(\ref{momcons}), 
and the relation of $\Theta^{ij}$ to $\Pi_c^{ij}$, Eq.~(\ref{thetapi}),  
are derived at once from the invariance of $I$ under general spatial coordinate 
transformations including the space translation and rotation. 
 We should also note that  
Eq.~(\ref{momcons}) can be further modified into a more physically familiar form 
of the momentum conservation law as shown 
in Eq.~(33) of Ref.~\cite{Sugama2013}. 

It is shown in Appendix~A that,  
reducing the field Lagrangian density given by Eq.~(\ref{Lf}) 
to the more simplified one defined in Eq.~(\ref{LVPfdens}) 
and regarding the vector potential in Eq.~(\ref{spL})
as a fixed time-independent field, 
the governing equations for the Vlasov-Poisson system 
can be obtained from the Eulerian variational principle 
in the same manner as shown for the Vlasov-Poisson-Amp\`{e}re system. 
   As pointed out by Qin {\it et al}.,~\cite{Qin2} 
when governing equations for a simplified system are obtained by
applying a certain approximation to a Lagrangian for another system, 
the exact energy and momentum conservation laws in the simplified system 
should be derived from the symmetry properties of the approximate Lagrangian and 
they generally disagree with those obtained by just 
making a similar approximation 
to the conservation laws in the original system. 
    The momentum balance and the energy conservation law 
in the Vlasov-Poisson system are 
derived in Appendices~A and B, respectively, 
where they are found to agree with those 
given by Qin {\it et al}.~\cite{Qin2}  

\section{DRIFT KINETIC SYSTEM}

In this section, the Eulerian variational principle is presented 
for the collisionless drift kinetic equation which governs 
the time evolution of the phase-space distribution function of 
guiding centers of charged particles in the strong magnetic field.
   The invariance of the drift kinetic system under 
an arbitrary spatial coordinate transformation is used to obtain   
the momentum balance as a three-dimensional vector equation 
in which the symmetric pressure tensor, the Lorentz force, and  
the magnetization current are properly included. 

\subsection{Eulerian variational principle for 
derivation of the collisionless drift kinetic equation}

We here start with defining 
the action integral for the drift kinetic system by 
\begin{equation}
\label{IDK}
I_{DK}  \equiv  \int_{t_1}^{t_2} dt \; L_{DK} 
\equiv \int_{t_1}^{t_2}  dt \int_V d^3 x\; {\cal L}_{DK} 
,
\end{equation}
where the Lagrangian density is written as 
\begin{equation}
\label{LDK}
{\cal L}_{DK} 
 \equiv 
\int d^3 v
\; F (x^i, v_\parallel,   \mu, \vartheta,  t) L_{GC}  
. 
\end{equation}
The guiding center position is represented 
in terms of the general spatial coordinates $(x^i)_{i=1,2,3}$,  
for which the metric tensor is given by $g_{ij}$. 
The velocity component of the guiding center along the magnetic field line, 
the magnetic moment, and the gyrophase angle are denoted by 
$v_\parallel$, $\mu$, and $\vartheta$, respectively. 
The integral with respect to the velocity space variables 
$(v_\parallel, \mu, \vartheta)$ is denoted by 
\begin{equation}
\int d^3 v
\equiv 
\int_{-\infty}^\infty d v_\parallel \int_0^\infty
 d\mu \oint d\vartheta
\end{equation}
and the Lagrangian for the single guiding center is given by  
\begin{eqnarray}
\label{Lgu}
& & 
L_{GC} \left[ v_\parallel,   \mu,  
u_x^i,  
u_\vartheta,  
\phi, A_i , \frac{\partial A_j}{\partial x^i}, g_{ij}  \right]
\nonumber \\ 
&  & 
= 
\left[ \frac{e}{c} A_j (x^n, t) + m v_\parallel b^i (x^n, t) g_{ij} (x^n) \right]  
u_x^j 
+ \frac{m c}{e} \mu u_\vartheta  
\nonumber \\  & & 
\hspace*{5mm}
\mbox{}
- \left[ \frac{1}{2} m v_\parallel^2 + \mu B (x^n, t) + e \phi (x^n, t) \right]
. 
\hspace*{5mm}
\end{eqnarray}
   Here, the unit vector parallel to the magnetic field is written as
\begin{equation}
b^i (x^n, t) 
= 
\frac{B^i (x^n, t)}{B (x^n, t)}
, 
\end{equation}
where the field strength is given by
\begin{equation}
B (x^n, t) 
=
\sqrt{ g_{ij} (x^n) B^i (x^n, t) B^j (x^n, t)}
, 
\end{equation}
and the $i$th contravariant component  $B^i$ of 
the magnetic field  is expressed in Eq.~(\ref{Bi}). 
The Lagrangian $L_{GC}$ shown in Eq.~(\ref{Lgu}) 
represents 
Littlejohn's guiding-center Lagrangian~\cite{Littlejohn}
written using the general spatial coordinates and the Eulerian picture. 

We now describe the particle's motion in the Lagrangian picture 
by representing the guiding center position coordinates,  
parallel velocity,  magnetic moment, and 
gyrophase at time $t$ as the functions  
$x_L^i (x_0^n, v_{\parallel 0}, \mu_0, \vartheta_0, t_0 ; t)$, 
$v_{\parallel L} (x_0^n, v_{\parallel 0}, \mu_0, \vartheta_0, t_0 ; t)$, 
$\mu_L (x_0^n, v_{\parallel 0}, \mu_0, \vartheta_0, t_0 ; t)$, 
and 
$\vartheta_L (x_0^n, v_{\parallel 0}, \mu_0, \vartheta_0, t_0 ; t)$, 
respectively, where $x_0^n$, $v_{\parallel 0}$, $\mu_0$, and 
$\vartheta_0$ denote their values at the initial time $t_0$. 
   Then, the distribution function 
$F (x^i, v_\parallel,   \mu, \vartheta,  t)$ at time $t$ 
is related to that at time $t_0$ by
\begin{eqnarray}
\label{Fgu}
& & F (x^i, v_\parallel,   \mu, \vartheta,  t)
\nonumber \\ 
& = & 
\int_{V_0} d^3 x_0 \int d^3 v_0
\; F (x_0^m, v_{\parallel 0},   \mu_0, \vartheta_0,  t_0)
\nonumber \\ & & 
\mbox{} \times
\delta^3 [x^i - x_L^i (x_0^m, v_{\parallel 0},   \mu_0, \vartheta_0, t_0; t) ]
\nonumber \\ & & 
\mbox{}
\times 
\delta [v_\parallel - v_{\parallel L} 
(x_0^m, v_{\parallel 0},   \mu_0, \vartheta_0, t_0; t) ]
\nonumber \\ & & 
\mbox{}
\times 
\delta [\mu - \mu_L 
(x_0^m, v_{\parallel 0},   \mu_0, \vartheta_0, t_0; t) ]
\nonumber \\ & & 
\mbox{}
\times 
\delta [\vartheta - \vartheta_L 
(x_0^m, v_{\parallel 0},   \mu_0, \vartheta_0, t_0; t) 
({\rm mod}\; 2\pi) ]
, 
\end{eqnarray}
where 
$
\int d^3 v_0
\equiv 
\int_{-\infty}^\infty d v_{\parallel 0} \int_0^\infty
 d\mu_0 \oint d\vartheta_0
$.
    In the Eulerian picture, 
the temporal change rates of the guiding center position, 
parallel velocity, magnetic moment, and gyrophase 
are denoted by the functions 
$u_x^i (x^m,  v_{\parallel},\mu, \vartheta,  t)$, 
$u_{v_\parallel} (x^m,  v_{\parallel},\mu, \vartheta,  t)$, 
$u_\mu (x^m,  v_{\parallel},\mu, \vartheta,  t)$, 
and $u_\vartheta (x^m,  v_{\parallel},\mu, \vartheta,  t)$, 
respectively, and they are related to those in the Lagrangian picture 
by 
\begin{eqnarray}
\label{udotxgu}
u_x^i (x_L^m,  v_{\parallel L},\mu_L, \vartheta_L,  t)
& = & 
\dot{x}_L^i (x_0^n, v_{\parallel 0}, \mu_0, \vartheta_0, t_0;  t), 
\nonumber \\ 
u_{v_\parallel} (x_L^m,  v_{\parallel L},\mu_L, \vartheta_L,  t)
& = & 
\dot{v}_{\parallel L} (x_0^n, v_{\parallel 0}, \mu_0, \vartheta_0, t_0;  t), 
\nonumber \\ 
u_\mu (x_L^m,  v_{\parallel L},\mu_L, \vartheta_L,  t)
& = & 
\dot{\mu}_L (x_0^n,  v_{\parallel 0}, \mu_0, \vartheta_0, t_0;  t), 
\nonumber \\ 
u_\vartheta (x_L^m,  v_{\parallel L},\mu_L, \vartheta_L,  t)
& = & 
\dot{\vartheta}_L (x_0^n, v_{\parallel 0}, \mu_0, \vartheta_0, t_0;  t), 
\hspace*{5mm}
\end{eqnarray}
   where 
$\dot{f}= \partial f (x_0^n, v_{\parallel 0}, \mu_0, \vartheta_0,  t)/ \partial t$ represents the time derivative of 
an arbitrary function $f (x_0^n, v_{\parallel 0}, \mu_0, \vartheta_0,  t)$ 
with $(x_0^n, v_{\parallel 0}, \mu_0, \vartheta_0)$ kept fixed. 
   It can be shown from Eqs.~(\ref{Fgu}) and (\ref{udotxgu}) 
that $F$ satisfies 
\begin{equation}
\label{DKE0}
\frac{\partial F}{\partial t} 
+ \frac{\partial}{\partial x^j}
( F u_x^j  )
+ \frac{\partial}{\partial v_\parallel}
( F u_{v_\parallel}  )
+ \frac{\partial}{\partial \mu}
( F u_\mu )
+ \frac{\partial}{\partial \vartheta}
( F u_\vartheta  )
= 0
. 
\end{equation}

The virtual displacement of the particle's trajectory in the 
$(x^i,  v_{\parallel},\mu, \vartheta)$ space is represented 
by the variations of the Lagrangian representations of the particle's 
motion as
$\delta x_L^i (x_0^n, v_{\parallel 0}, \mu_0, \vartheta_0, t_0 ; t)$, 
$\delta  v_{\parallel L} (x_0^n, v_{\parallel 0}, \mu_0, \vartheta_0, t_0 ; t)$, 
$\delta \mu_L (x_0^n, v_{\parallel 0}, \mu_0, \vartheta_0, t_0 ; t)$, 
and 
$\delta \vartheta_L (x_0^n, v_{\parallel 0}, \mu_0, \vartheta_0, t_0 ; t)$. 
   The variations in the guiding center position,  parallel velocity,  
magnetic moment, and gyrophase are represented in the Eulerian picture by
$\delta x_E^i (x^n, v_\parallel, \mu, \vartheta, t)$, 
$\delta  v_{\parallel E} (x^n, v_\parallel, \mu, \vartheta, t)$, 
$\delta \mu_E (x^n, v_\parallel, \mu, \vartheta, t)$, 
and 
$\delta \vartheta_E (x^n, v_\parallel, \mu, \vartheta, t)$, 
which are related to those in the Lagrangian picture by
\begin{eqnarray}
\label{dxvmt}
\delta  x_E^i 
(x_L^m,  v_{\parallel L},\mu_L, \vartheta_L,  t)
& = & 
\delta  x_L^i (x_0^n, v_{\parallel 0}, \mu_0, \vartheta_0, t_0 ; t),
\nonumber \\
\delta  v_{\parallel E}
(x_L^m,  v_{\parallel L},\mu_L, \vartheta_L,  t)
& = & 
\delta  v_{\parallel L} (x_0^n, v_{\parallel 0}, \mu_0, \vartheta_0, t_0 ; t),
\nonumber \\
\delta  \mu_E
(x_L^m,  v_{\parallel L},\mu_L, \vartheta_L,  t)
& = & 
\delta  \mu_L (x_0^n, v_{\parallel 0}, \mu_0, \vartheta_0, t_0 ; t),
\nonumber \\
\delta  \vartheta_E
(x_L^m,  v_{\parallel L},\mu_L, \vartheta_L,  t)
& = & 
\delta  \vartheta_L (x_0^n, v_{\parallel 0}, \mu_0, \vartheta_0, t_0 ; t). 
\hspace*{6mm}
\end{eqnarray}
   Using Eqs.~(\ref{udotxgu}) and (\ref{dxvmt}), 
the variations in the functional forms of $u_x^i$, 
$u_{v_\parallel}$, $u_\mu$, and $u_\vartheta$ 
due to the virtual displacement of the particle's trajectory are 
given by
\begin{eqnarray}
\label{duvmt}
\delta u_x^i
& = & 
\left( \frac{\partial }{\partial t}
+ u_x^j \frac{\partial }{\partial x^j}
+ u_{v_\parallel} \frac{\partial }{\partial v_\parallel}
+ u_\mu \frac{\partial }{\partial \mu}
+ u_\vartheta \frac{\partial }{\partial \vartheta}
\right) \delta x_E^i
\nonumber \\ 
& & \mbox{}
- \left( 
\delta x_E^j \frac{\partial }{\partial x^j}
+ \delta v_{\parallel E} \frac{\partial }{\partial v_\parallel}
+  \delta \mu_E \frac{\partial }{\partial \mu}
+ \delta \vartheta_E \frac{\partial }{\partial \vartheta}
\right) u_x^i
,
\nonumber \\ 
\delta u_{v_\parallel}
& = & 
\left( \frac{\partial }{\partial t}
+ u_x^j \frac{\partial }{\partial x^j}
+ u_{v_\parallel} \frac{\partial }{\partial v_\parallel}
+ u_\mu \frac{\partial }{\partial \mu}
+ u_\vartheta \frac{\partial }{\partial \vartheta}
\right) \delta v_{\parallel E}
\nonumber \\ 
& & \mbox{}
- \left( 
\delta x_E^j \frac{\partial }{\partial x^j}
+ \delta v_{\parallel E} \frac{\partial }{\partial v_\parallel}
+  \delta \mu_E \frac{\partial }{\partial \mu}
+ \delta \vartheta_E \frac{\partial }{\partial \vartheta}
\right) u_{v_\parallel}
,
\nonumber \\ 
\delta u_\mu
& = & 
\left( \frac{\partial }{\partial t}
+ u_x^j \frac{\partial }{\partial x^j}
+ u_{v_\parallel} \frac{\partial }{\partial v_\parallel}
+ u_\mu \frac{\partial }{\partial \mu}
+ u_\vartheta \frac{\partial }{\partial \vartheta}
\right) \delta \mu_E
\nonumber \\ 
& & \mbox{}
- \left( 
\delta x_E^j \frac{\partial }{\partial x^j}
+ \delta v_{\parallel E} \frac{\partial }{\partial v_\parallel}
+  \delta \mu_E \frac{\partial }{\partial \mu}
+ \delta \vartheta_E \frac{\partial }{\partial \vartheta}
\right) u_\mu
,
\nonumber \\ 
\delta u_\vartheta
& = & 
\left( \frac{\partial }{\partial t}
+ u_x^j \frac{\partial }{\partial x^j}
+ u_{v_\parallel} \frac{\partial }{\partial v_\parallel}
+ u_\mu \frac{\partial }{\partial \mu}
+ u_\vartheta \frac{\partial }{\partial \vartheta}
\right) \delta \vartheta_E
\nonumber \\ 
& & \mbox{}
- \left( 
\delta x_E^j \frac{\partial }{\partial x^j}
+ \delta v_{\parallel E} \frac{\partial }{\partial v_\parallel}
+  \delta \mu_E \frac{\partial }{\partial \mu}
+ \delta \vartheta_E \frac{\partial }{\partial \vartheta}
\right) u_\vartheta
. 
\nonumber \\ 
& & \mbox{}
\end{eqnarray}
The variation in the distribution function due to the virtual displacement of 
the particle's trajectory is written by 
using Eqs.~(\ref{Fgu}) and (\ref{dxvmt}) 
as
\begin{equation}
\label{dFDK}
\delta F
=
- \frac{\partial}{\partial x^j}
( F \delta x_E^j  )
- \frac{\partial}{\partial v_\parallel}
( F \delta v_{\parallel E})
- \frac{\partial}{\partial \mu}
( F \delta \mu_E)
- \frac{\partial}{\partial \vartheta}
( F \delta \vartheta_E)
. 
\end{equation}

Using Eqs.~(\ref{DKE0}), (\ref{duvmt}), and (\ref{dFDK}), 
we find that the variation in 
the action integral $I_{DK}$ due to 
the virtual displacement of particle's trajectory 
is written as 
\begin{eqnarray}
\label{dIDK}
\hspace*{-14mm}
\delta I_{DK}
& = & 
\int_{t_1}^{t_2}  dt \int_V d^3 x 
\int d^3 v
\left[
F  \left\{ 
\left( \frac{\partial L_{GC}}{\partial x^i} \right)_u
\right. \right.
\nonumber \\
& & 
\hspace*{-6mm}
\left. 
\mbox{} 
-    \frac{d}{d t } 
\left( \frac{\partial L_{GC}}{\partial u_x^i} \right)
\right\} \delta x_E^i 
+ F
\left( \frac{\partial L_{GC}}{\partial v_\parallel} \right)_u \delta v_{\parallel E}
\nonumber \\
& & 
\hspace*{-6mm}
\mbox{} + F
\left( \frac{\partial L_{GC}}{\partial \mu } \right)_u \delta \mu_E
+ 
F  \left\{ 
\left( \frac{\partial L_{GC}}{\partial \vartheta} \right)_u
\right. \nonumber \\ & & 
\hspace*{-6mm}
\left. 
\mbox{}
-    \frac{d}{d t } 
\left( \frac{\partial L_{GC}}{\partial u_\vartheta} \right)
\right\} \delta \vartheta_E
\nonumber \\
& & 
\hspace*{-6mm}
\mbox{}
+
\frac{\partial}{\partial t } \left\{
F \left( \frac{\partial L_{GC}}{\partial u_x^i }  \delta x_E^i
+ \frac{\partial L_{GC}}{\partial u_\vartheta}  \delta \vartheta_E
\right)
\right\}
\nonumber \\
& & 
\hspace*{-6mm}
\mbox{}
+
\frac{\partial}{\partial x^j} \left\{
F  u_x^j \left( \frac{\partial L_{GC}}{\partial u_x^i }  \delta x_E^i
+ \frac{\partial L_{GC}}{\partial u_\vartheta}  \delta \vartheta_E
\right)
\right\}
\nonumber \\
& & 
\hspace*{-6mm}
\mbox{}
+
\frac{\partial}{\partial v_\parallel } \left\{
F  u_{v_\parallel} \left( \frac{\partial L_{GC}}{\partial u_x^i }  \delta x_E^i
+ \frac{\partial L_{GC}}{\partial u_\vartheta}  \delta \vartheta_E
\right)
\right\}
\nonumber \\
& & 
\hspace*{-6mm}
\mbox{}
+
\frac{\partial}{\partial \mu} \left\{
F  u_\mu \left( \frac{\partial L_{GC}}{\partial u_x^i }  \delta x_E^i
+ \frac{\partial L_{GC}}{\partial u_\vartheta}  \delta \vartheta_E
\right)
\right\}
\nonumber \\
& & 
\hspace*{-6mm}
\left. 
\mbox{}
+
\frac{\partial}{\partial \vartheta} \left\{
F  u_\vartheta \left( \frac{\partial L_{GC}}{\partial u_x^i }  \delta x_E^i
+ \frac{\partial L_{GC}}{\partial u_\vartheta}  \delta \vartheta_E
\right)
\right\}
\right]
, 
\end{eqnarray}
where 
$( \partial L_{GC}/\partial x^i )_u$, 
$( \partial L_{GC}/\partial v_\parallel )_u$, 
$( \partial L_{GC}/\partial \mu )_u$, 
 and 
$( \partial L_{GC}/\partial \vartheta )_u$ denote  
the derivatives of $L_{GC}$ in $x^i$, $v_\parallel$, $\mu$, 
and $\vartheta$, respectively, 
with $(u_x^i, u_\vartheta)$ kept fixed in $L_{GC}$, and 
the time derivative along the particle's trajectory 
is represented by 
\begin{equation}
\label{ddt2}
\frac{d}{dt}
 \equiv  
\frac{\partial}{\partial t } 
+  u_x^k \frac{\partial}{\partial x^k} 
+ u_{v_\parallel} \frac{\partial}{\partial v_\parallel} 
+  u_\mu \frac{\partial}{\partial \mu} 
+  u_\vartheta \frac{\partial}{\partial \vartheta} 
. 
\end{equation}

We now use the Eulerian variational principle which implies that 
the collisionless drift kinetic equation for the distribution function $F$ 
can be derived from the condition that $\delta I_{DK} = 0$ 
for arbitrary variations 
$\delta x_E^i$, $\delta v_{\parallel E}$, $\delta \mu_E$, and 
$\delta \vartheta_E$ which vanish on the boundaries of the integral region. 
  We first use 
$\delta I_{DK}/\delta x_E^i =0$
to obtain 
\begin{equation}
\label{dpidt}
\frac{d}{dt} p_i
=
\left( \frac{\partial L_{GC}}{\partial x^i} \right)_u
, 
\end{equation}
where $p_i$ represents the covariant vector component 
of the canonical momentum defined by
\begin{equation}
p_i
\equiv 
\frac{\partial L_{GC}}{\partial u_x^i} 
= 
\frac{e}{c} A_i (x^n, t) + m v_\parallel b_i (x^n, t)
\equiv \frac{e}{c} A^*_i (x^n, v_\parallel, t) 
. 
\end{equation}
We should note that 
the distribution function $F$ is included as a factor 
in  $\delta I_{DK}/\delta x_E^i =0$ 
although it is omitted from Eq.~(\ref{dpidt}) for simplicity 
in the same way as done in Sec.~II.A. 
This omission of $F$ is also done in the other equations obtained below from 
$\delta I_{DK} =0$ although it does not make a difference in deriving 
the resultant collisionless drift kinetic equation in Eq.~(\ref{DKE}). 
   We can rewrite  Eq.~(\ref{dpidt}) as 
\begin{equation}
\label{mupb}
m u_{v_\parallel} b_i  =
 e \left( E^*_i + \frac{1}{c} \sqrt{g} \epsilon_{ijk} u_x^j 
B^{*k} \right)
- \mu \frac{\partial B}{\partial x^i}
, 
\end{equation}
where the modified electric and magnetic fields are defined by 
\begin{equation}
E^*_i  \equiv
- \frac{\partial \phi}{\partial x^i}
-   \frac{1}{c}
\frac{\partial  A^*_i}{\partial t}
,
\end{equation}
and 
\begin{equation}
B^{*i}  \equiv
\frac{\epsilon^{ijk}}{\sqrt{g}} \frac{\partial A^*_k}{\partial x^j}
,
\end{equation}
respectively. 

Next, 
$\delta I_{DK}/\delta v_{\parallel E} =0$
is used to obtain 
\begin{equation}
\label{dLdv}
\left( \frac{\partial L_{GC}}{\partial v_\parallel} \right)_u
= m \left( 
u_x^i b_i  - v_\parallel 
\right)
= 0 
, 
\end{equation}
from which we have 
\begin{equation}
\label{ubvp}
u_x^i b_i  
=
 v_\parallel 
.
\end{equation}
   Furthermore,  
$\delta I_{DK}/\delta \mu_E =0$
and $\delta I_{DK}/\delta \vartheta_E =0$ 
yield 
\begin{equation}
\label{dIdmu}
\left( \frac{\partial L_{GC}}{\partial \mu} \right)_u
= 
\frac{m c}{e} u_\vartheta - B
= 0 
,
\end{equation}
and 
\begin{equation}
\label{dIdth}
\frac{d}{d t } 
\left( \frac{\partial L_{GC}}{\partial u_\vartheta} \right)
= 
\frac{m c}{e} u_\mu
=
\left( \frac{\partial L_{GC}}{\partial \vartheta} \right)_u
=
0  
, 
\end{equation}
respectively. 

Equations~(\ref{mupb}), (\ref{ubvp}), (\ref{dIdmu}), and 
(\ref{dIdth}) are rewritten as 
\begin{equation}
\label{uxieq}
u_x^i
= 
\frac{1}{B^*_\parallel}
\left[ v_\parallel B^{*i}
+
c \frac{\epsilon^{ijk}}{\sqrt{g}} b_j
\left( 
 \frac{\mu}{e} \frac{\partial B}{\partial x^k}
- E^*_k 
\right)
\right]
,
\end{equation}
\begin{equation}
\label{uvpeq}
m u_{v_\parallel} 
= 
\frac{B^{*i}}{B^*_\parallel}
 \left(
e  E^*_i 
- \mu \frac{\partial B}{\partial x^i}
\right)
,
\end{equation}
\begin{equation}
\label{umueq}
u_\mu
= 
0
,
\end{equation}
   and 
\begin{equation}
\label{uvteq}
u_\vartheta 
= 
\frac{e B}{m c} 
\equiv \Omega
, 
\end{equation}
 where 
\begin{equation}
B^*_\parallel
\equiv
B^{*i} b_i
. 
\end{equation}
Equations~(\ref{uxieq}) and (\ref{uvpeq}) are obtained by 
taking the vector and scalar products between the magnetic 
field and Eq.~(\ref{mupb}), respectively. 
Also, using Eq.~(\ref{ddt2}), we can write 
\begin{equation}
\label{ddt3}
\left( u_x^i,  u_{v_\parallel}, u_\mu, u_\vartheta \right)
= 
\left( \frac{d x^i}{dt}, \frac{dv_\parallel}{dt}, 
\frac{d \mu}{dt}, \frac{d \vartheta}{dt} \right)
. 
\end{equation}
Then, with the help of Eq.~(\ref{ddt3}), 
it is clearly confirmed that Eqs.~(\ref{uxieq})--(\ref{uvteq}) 
represent the same guiding center motion equations as 
derived by Littlejohn from the guiding center Lagrangian. 
We can verify that the right-hand sides of Eqs.~(\ref{uxieq})--(\ref{uvteq}) 
are all independent of $\vartheta$ and that the magnetic moment 
$\mu$ is an invariant of motion as seen from Eq.~(\ref{umueq}). 

Substituting Eqs.~(\ref{uxieq})--(\ref{uvteq}) into 
Eqs.~(\ref{DKE0}) and taking its average with respect to 
the gyrophase $\vartheta$, 
the collisionless drift kinetic equation is derived as 
\begin{eqnarray}
\label{DKE}
& & 
\frac{\partial \overline{F}}{\partial t} 
+ \frac{\partial}{\partial x^i}
\left( \overline{F} \frac{1}{B^*_\parallel}
\left[ v_\parallel B^{*i}
+
c \frac{\epsilon^{ijk}}{\sqrt{g}} b_j
\left( 
 \frac{\mu}{e} \frac{\partial B}{\partial x^k}
- E^*_k 
\right)
\right]
 \right)
\nonumber \\ 
& & 
+ \frac{\partial}{\partial v_\parallel}
\left( 
\overline{F} 
\frac{B^{*i}}{m B^*_\parallel}
 \left(
e  E^*_i 
- \mu \frac{\partial B}{\partial x^i}
\right)
 \right)
 = 0
, 
\end{eqnarray}
where $\overline{F}$ denotes the gyrophase-averaged 
distribution function, 
\begin{equation}
\overline{F}
\equiv 
\oint \frac{d\vartheta}{2\pi} F
. 
\end{equation}

\subsection{Transformation of spatial coordinates}

Here, in the same way as in Sec.~II.B, 
the infinitesimal transformation of the spatial coordinates is 
given by Eq.~(\ref{xprime}) and 
the infinitesimal variation $\xi^i$ in the spatial coordinate $x^i$ 
is again regarded as an arbitrary function of only the spatial coordinates. 
However, it should be noted that 
the other variables $(v_\parallel, \mu, \vartheta)$ are independent of 
the choice of the spatial coordinates because 
they are defined from the relation of the velocity vector to the direction of 
the local magnetic field. 
This is in contrast to the case of Sec.~II.B where  
the velocity components  $(v^i)_{i=1,2,3}$ are transformed as 
the contravariant vector components under 
the spatial coordinate transformation.

The spatial coordinate transformation
 given by Eq.~(\ref{xprime}) changes 
the Lagrangian representation of the guiding center position as
\begin{eqnarray}
\label{xxi}
& &
\hspace*{-8mm}
 x'^i_L (x'^n_0, v_{\parallel 0}, \mu_0, \vartheta_0, t_0 ; t)
=
x^i_L (x_0^n, v_{\parallel 0}, \mu_0, \vartheta_0, t_0 ; t)
\nonumber \\ & & \mbox{}
\hspace*{25mm}
+ \xi^i ( x^m_L (x_0^n, v_{\parallel 0}, \mu_0, \vartheta_0, t_0 ; t) )
,
\hspace*{5mm}
\end{eqnarray}
where 
$
x'^i_0 
=
x^i_0 + \xi^i ( x^n_0  ) 
$
represents the guiding center position at time $t_0$ 
in the transformed spatial coordinates. 
   The distribution function is written in the transformed coordinates as 
\begin{eqnarray}
\label{FPDK}
& & F' (x'^i, v_\parallel,   \mu, \vartheta,  t)
\nonumber \\ 
& = & 
\int_{V'_0} d^3 x'_0 
\int d^3 v_0
\; F' (x'^m_0, v_{\parallel 0},   \mu_0, \vartheta_0,  t_0)
\nonumber \\ & & 
\mbox{} \times
\delta^3 [x'^i - x'^i_L (x'^m_0, v_{\parallel 0},   \mu_0, \vartheta_0, t_0; t) ]
\nonumber \\ & & 
\mbox{}
\times 
\delta [v_\parallel - v_{\parallel L} 
(x'^m_0, v_{\parallel 0},   \mu_0, \vartheta_0, t_0; t) ]
\nonumber \\ & & 
\mbox{}
\times 
\delta [\mu - \mu_L 
(x'^m_0, v_{\parallel 0},   \mu_0, \vartheta_0, t_0; t) ]
\nonumber \\ & & 
\mbox{}
\times 
\delta [\vartheta - \vartheta_L 
(x'^m_0, v_{\parallel 0},   \mu_0, \vartheta_0, t_0; t) 
({\rm mod}\; 2\pi) ]
.
\hspace*{5mm}
\end{eqnarray}
   The initial distribution functions
$F' (x'^n_0, v_{\parallel 0},   \mu_0, \vartheta_0, t_0)$ 
and 
$F (x^n_0, v_{\parallel 0},   \mu_0, \vartheta_0, t_0)$  
 in the transformed and original coordinate systems are 
related to each other by 
$
F' (x'^n_0, v_{\parallel 0},   \mu_0, \vartheta_0, t_0) 
d^3 x'_0 
= 
F (x^n_0, v_{\parallel 0},   \mu_0, \vartheta_0, t_0) 
d^3 x_0 
, 
$
which is rewritten as 
\begin{equation}
\label{FFDK}
F' (x'^n_0, v_{\parallel 0},   \mu_0, \vartheta_0, t_0) 
= 
F (x^n_0, v_{\parallel 0},   \mu_0, \vartheta_0, t_0) 
\left[
\det \left( \frac{\partial x'^i_0}{\partial x^j_0} \right)
\right]^{-1}
. 
\end{equation}
The variation $\overline{\delta} F$ in the functional form of 
the distribution function $F$ 
due to the spatial coordinate transformation 
is defined by 
\begin{equation}
\label{dFDK2}
F' (x^i, v_\parallel,   \mu, \vartheta,  t)
= F (x^i, v_\parallel,   \mu, \vartheta,  t)
+ \overline{\delta} F (x^i, v_\parallel,   \mu, \vartheta,  t)
.
\end{equation}
Then, it is shown by using Eqs.~(\ref{xxi}), (\ref{FPDK}), (\ref{FFDK}), 
and (\ref{dFDK2}) that 
$\overline{\delta} F$ can be represented by 
\begin{equation}
\label{dFDK3}
\overline{\delta} F
= 
 - \frac{\partial}{\partial x^j}
( F \xi^j  )
. 
\end{equation}

In the same way as seen in Sec.~II.B,  
the variations 
 in the functional forms of $u_x^i$, $u_{v_\parallel}$, 
$u_\mu$, $u_\vartheta$, $\phi$, $A_i$, and $g_{ij}$ 
due to 
the spatial coordinate transformation
are denoted by 
$\overline{\delta} u_x^i$, 
$\overline{\delta} u_{v_\parallel}$, 
$\overline{\delta} u_\mu$, 
$\overline{\delta} u_\vartheta$, 
$\overline{\delta} \phi$, 
$\overline{\delta} A_i$, and $\overline{\delta} g_{ij}$, 
respectively. 
They can be represented by 
using the Lie derivative $L_\xi$ as 
\begin{eqnarray}
\label{dupAg}
 \overline{\delta} u_x^i
& = & -  L_\xi  u_x^i  
\equiv  
- \xi^j  \frac{\partial u_x^i}{\partial x^j} 
+  u_x^j \frac{\partial \xi^i}{\partial x^j}
,
\nonumber \\ 
\overline{\delta} u_{v_\parallel}
& = & -  L_\xi  u_{v_\parallel}  
\equiv  
- \xi^j  \frac{\partial u_{v_\parallel}}{\partial x^j} 
,
\nonumber \\ 
\overline{\delta} u_\mu
& = & -  L_\xi  u_\mu  
\equiv  
- \xi^j  \frac{\partial u_\mu}{\partial x^j} 
,
\nonumber \\ 
\overline{\delta} u_\vartheta
& = & -  L_\xi  u_\vartheta  
\equiv  
- \xi^j  \frac{\partial u_\vartheta}{\partial x^j} 
.
\end{eqnarray}
The expressions of 
$\overline{\delta} \phi$, 
$\overline{\delta} A_i$, and  
$\overline{\delta} g_{ij}$ in terms of the Lie derivative are shown in 
Eqs.~(\ref{deltaphi})
(\ref{deltaA}), and (\ref{dEgg}), respectively.  

\subsection{Derivation of the momentum balance equation}

We can use Eqs.~(\ref{IDK})--(\ref{Lgu}), (\ref{dFDK3}), and (\ref{dupAg})
to derive the action integral $I'_{DK} = I_{DK} + \overline{\delta}  I_{DK}$ 
in the transformed spatial coordinates. 
Here, the variation $\overline{\delta}  I_{DK}$ in the action integral due to the 
spatial coordinate transformation is written as 
\begin{equation}
\label{dIDK2}
\overline{\delta}  I_{DK}  
=
\int_{t_1}^{t_2}  dt \int_V d^3 x 
 \left[
\xi_j  J_{DK}^j 
+
\frac{\partial }{\partial x^i} \left(\xi_j T_{DK}^{ij}  \right)
\right]
,
\end{equation}
where the vector density $J_{DK}^j$ and 
the tensor density $T_{DK}^{ij}$ 
are defined by 
\begin{eqnarray}
\label{Jj0}
J_{DK}^j
& \equiv & 
g^{jk}
\int d^3 v \left[
\frac{\partial }{\partial t} ( F m v_\parallel b_k )
 + \sqrt{g} \epsilon_{klm} B^l 
\right. 
\nonumber \\ & & 
 \mbox{}
\times
\left\{ \frac{e}{c} F u_x^m
- \frac{\partial }{\partial x^n} 
\left( F \frac{\partial L_{GC}}{\partial (\partial A_m/ \partial x^n)} 
\right) \right\}
\nonumber \\ & & \mbox{}
\left. 
- e F E_k 
+ 2  \nabla_l
\left( F g_{km} \frac{\partial L_{GC}}{\partial g_{lm}} 
\right)
\right]
,
\end{eqnarray}
and 
\begin{eqnarray}
\label{Tij0}
T_{DK}^{ij}
& \equiv & 
g^{jk}
\int d^3 v \,
F \left[
u_x^i 
\frac{\partial L_{GC}}{\partial u_x^k}
-2 g_{kl} \frac{\partial L_{GC}}{\partial g_{il}}
- \frac{\partial L_{GC}}{\partial A_i} A_k
\right. 
\nonumber \\ & &
\left.  \mbox{}
+ \frac{\partial L_{GC}}{\partial (\partial A_l/\partial x^i)}
\left( 
 \frac{\partial A_k}{\partial x^l}
-  \frac{\partial A_l}{\partial x^k}
\right) \right]
,
\end{eqnarray}
respectively. 
 In deriving Eq.~(\ref{dIDK2}), we also need to use
Eqs.~(\ref{dpidt}), (\ref{dLdv}),  (\ref{dIdmu}), and 
(\ref{dIdth}) obtained from the 
Eulerian variational principle in Sec.~III.A. 

Note that the action integral $I_{DK}$ is invariant under the spatial coordinate transformation and 
accordingly $\overline{\delta} I_{DK}$ shown in Eq.~(\ref{dIDK2}) vanish for 
any $\xi_j$. 
 Then, the integrands at the interior and boundary points 
on the right-hand side of Eq.~(\ref{dIDK2}) must 
vanish separately. 
Thus, we obtain $J_{DK}^i = 0$ and $T_{DK}^{ij} = 0$. 
   Substituting Eq.~(\ref{Lgu}) into Eq.~(\ref{Jj0}), 
the momentum balance equation is obtained from 
$J_{DK}^j  = 0$ as 
\begin{equation}
\label{mombal}
\frac{\partial}{\partial t} 
\left(  m N V_{g \parallel} b^j \right)
= e N
\left(  E^j + \frac{1}{c} 
\frac{\epsilon^{jkl}}{\sqrt{g}} V_k B_l  \right) 
- \nabla_i P^{ij}
,
\end{equation}
where 
\begin{equation}
\label{N}
 N
\equiv 
\int d^3 v \, F
,
\hspace*{5mm}
 N V_{g \parallel}
\equiv 
\int d^3 v \, F v_\parallel
, 
\end{equation}
and
\begin{eqnarray}
\label{NV}
 N V^k
& \equiv & 
\int d^3 v \, F u_x^k
+ \frac{c}{e} \epsilon^{kij}
\frac{\partial }{\partial x^i}
\left(
\int d^3 v \, 
\frac{F}{\sqrt{g}}
\right.
\nonumber \\ & & 
\left.
\times \left[
- \mu b_j +\frac{m v_\parallel}{B}
\left\{
(u_x)_j - (u_x)_l b^l b_j
\right\}
\right]
\right)
\hspace*{5mm}
\end{eqnarray}
are used. 
We see that the inertia term in the momentum balance equation, Eq.~(\ref{mombal}), 
contains only the parallel momentum component while
 the electric current $e NV^k$ in the Lorentz force term 
consists of the guiding-center current and  
the magnetization current~\cite{H&M} as shown in 
Eq.~(\ref{NV}). 
   The symmetric pressure tensor density $P^{ij}$ 
on the right-hand side of Eq.~(\ref{mombal}) 
is defined by 
\begin{equation}
\label{Pij}
P^{ij}
 \equiv 
2 \int d^3 v \; F \frac{\partial L_{GC}}{\partial g_{ij}}
=
P_{\rm CGL}^{ij}
+ \pi_\land^{ij}
, 
\end{equation}
where $P_{\rm CGL}^{ij}$ is given in 
the Chew-Goldberger-Low (CGL) form,~\cite{H&S}  
\begin{equation}
\label{CGL0}
P_{\rm CGL}^{ij}
=
\int d^3 v \, F
[ m v_\parallel^2 b^i b^j + \mu B ( g^{ij} - b^i b^j ) ]
, 
\end{equation}
and $\pi_\land^{ij}$ is the non-CGL part written as  
\begin{equation}
\label{non-CGL0}
\pi_\land^{ij}
\equiv 
 \int d^3 v \, F
m v_\parallel 
[ b^i ( u_x )_\perp^j  + ( u_x )_\perp^i b^j ]
.
\end{equation}
Here, the perpendicular component of the 
guiding center velocity is represented by 
$
( u_x )_\perp^i 
\equiv 
 u_x^i  - u_x^k b_k b^i
$. 
The symmetric pressure tensor given by Eq.~(\ref{Pij}) 
with Eqs.~(\ref{CGL0}) and (\ref{non-CGL0}) 
agrees with that given by Eq.~(19) in Ref.~\cite{Sugama2016}. 
The CGL pressure tensor shown in Eq.~(\ref{CGL0}) contains 
the scalar (or isotropic) part, which represents background pressure, 
and the anisotropic part, the magnitude of which is 
considered to be smaller than the background pressure 
by the factor $\sim \rho/L$ in the neoclassical transport theory. 
   Here, $\rho$ and $L$ represent 
the gyroradius and the equilibrium gradient scale length, respectively. 
  On the other hand, the magnitude of the non-CGL pressure tensor 
defined in Eq.~(\ref{non-CGL0})
is regarded as $\sim (\rho/L)^2$. 

We next substitute Eq.~(\ref{Lgu}) into Eq.~(\ref{Tij0}).  
  Then, the other condition, $T_{DK}^{ij} = 0$, derived from  
putting $\delta I_{DK}  = 0$ in Eq.~(\ref{dIDK2})
can be written as 
\begin{equation}
P^{ij}
= 
P_c^{ij}
+ D^{ij}
, 
\end{equation}
where $P^{ij}$ is the symmetric pressure tensor given 
by Eq.~(\ref{Pij})
and 
$P_c^{ij}$ is define by  
\begin{equation}
P_c^{ij}
\equiv 
g^{jk}
\int d^3 v \, F
u_x^i 
\frac{\partial L_{GC}}{\partial u_x^k}
= 
\int d^3 v \, F
u_x^i \left( m v_\parallel b^j + \frac{e}{c} A^j \right)
. 
\end{equation}
Here, $P_c^{ij}$ is an asymmetric tensor density representing 
the transport of the canonical momentum. 
The difference $D^{ij}$ between $P^{ij}$ and $P_c^{ij}$ 
is written as 
\begin{eqnarray}
& & 
\hspace*{-3mm}
D^{ij}
 \equiv  
g^{jk} 
\int d^3 v \, F \left[
- \frac{\partial L_{GC}}{\partial A_i} A_k
+ \frac{\partial L_{GC}}{\partial (\partial A_l/\partial x^i)}
\left( 
 \frac{\partial A_k}{\partial x^l}
-  \frac{\partial A_l}{\partial x^k}
\right) \right]
\nonumber \\ 
& & 
 =
\int d^3 v \, F
\left[ - \frac{e}{c} u_x^i  A^j 
+ m v_\parallel b^i ( u_x )_\perp^j 
+ \mu B ( g^{ij} - b^i b^j )
\right]
. 
\end{eqnarray}

\section{DRIFT KINETIC SYSTEM WITH SELF-CONSISTENT FIELDS}

In this section, not only the drift kinetic equations but also 
the equations for self-consistently generated electromagnetic fields 
are treated as constituents of the governing equations of the 
extended drift kinetic system. 
   The Eulerian variational principle is used to present all the governing equations 
and to derive the momentum conservation law satisfied by them.   
   The energy conservation law in the extended drift kinetic system is 
derived in Appendix~C where the energy balance in the drift kinetic system 
considered in Sec.~III is also obtained. 

\subsection{Quasineutrality and Amp\`{e}re's law combined with drift kinetic equations}

We here combine the quasineutrality condition and Amp\`{e}re's law 
with the drift kinetic equations in order to simultaneously determine 
the electromagnetic fields and the distribution functions for all particle species. 
  The action integral $I_{DKF}$ for deriving all the governing equations 
is written as 
\begin{equation}
\label{IDKF}
I_{DKF}  \equiv  \int_{t_1}^{t_2} dt \; L_{DKF} 
\equiv \int_{t_1}^{t_2}  dt \int_V d^3 x\; {\cal L}_{DKF} 
,
\end{equation}
where the Lagrangian density ${\cal L}_{DKF}$  is given by 
\begin{equation}
\label{LDKF}
{\cal L}_{DKF} 
 \equiv 
\sum_a
\int d^3 v
\; F_a  L_{GCa}  
- \frac{\sqrt{g}}{8 \pi} B^2
.  
\end{equation}
Here, the subscript $a$ represents the particle species. 
It is seen from Eq.~(\ref{LDKF}) 
that ${\cal L}_{DKF}$ contains
the summation of the drift kinetic Lagrangian densities [see Eq.~(\ref{LDK})] 
over all species 
 and 
the magnetic energy density with the minus sign. 

We now virtually let the trajectories of particles for all species, 
the electrostatic potential, and the vector potential vary infinitesimally. 
  Then, the resulting variation $\delta I_{DKF}$ in 
the action integral $I_{DKF}$ is expressed as  
\begin{eqnarray}
\delta I_{DKF}
& = & 
\sum_a
\int_{t_1}^{t_2}  dt \int_V d^3 x
\int d^3 v
\; F_a
 \left[
\left\{ 
\left( \frac{\partial L_{GCa}}{\partial x^i} \right)_u
\right. \right. 
\nonumber \\
& & 
\left. \left. 
\hspace*{-17mm}
-
\left(\frac{d}{d t}\right)_a
\left( \frac{\partial L_{GCa}}{\partial u_{ax}^i} \right)
\right\} \delta x_{aE}^i 
+ 
\left( \frac{\partial L_{GCa}}{\partial v_\parallel} \right)_u \delta v_{\parallel a E}
\right. 
\nonumber \\
& & 
\hspace*{-17mm}
\mbox{} + 
\left( \frac{\partial L_{GC}}{\partial \mu } \right)_u \delta \mu_{aE}
+ 
  \left\{ 
\left( \frac{\partial L_{GCa}}{\partial \vartheta} \right)_u
-    \left(\frac{d}{d t}\right)_a 
\left( \frac{\partial L_{GCa}}{\partial u_{a\vartheta}} \right)
\right\} 
\nonumber \\
& & 
\left. 
\hspace*{-17mm}
\times \delta \vartheta_{aE}
\right]
+ 
\int_{t_1}^{t_2} dt \int d^3 x 
\left[
- \delta \phi 
\sum_a e_a 
\int d^3 v  \; F_a 
\right. 
\nonumber \\
& &
\left. 
 \hspace*{-17mm} \mbox{}
+ \delta A_i \left\{
\sum_a \frac{e_a}{c} 
\int d^3 v \; F_a  u_{ax}^i
- \frac{\epsilon^{ijk}}{4\pi}  \frac{\partial }{\partial x^j} 
\left(
B_k - 4 \pi M_k
\right)
\right\}
\right]
\nonumber \\
& & \hspace*{-17mm}
\mbox{}
+ \delta I_{DKFb}
, 
\end{eqnarray}
    where 
$(d/dt)_a$ denotes the time derivative along the trajectory 
of the particle of species $a$ [see Eq.~(\ref{ddt2})], 
$\delta I_{DKFb}$ represents the part which is written as 
the boundary integrals,  
and $M_k = g_{kl} M^l$ is the $k$th covariant component 
of the magnetization vector. 
    The $k$th contravariant component of 
the magnetization vector is defined by 
\begin{equation}
\label{Mk}
M^k 
 \equiv 
\frac{1}{\sqrt{g}}
\sum_a  
\int d^3 v
 \; F_a 
\left(
- \mu b^k
+ \frac{m_a v_\parallel}{B} (u_{a x})_\perp^k
\right)
. 
\end{equation}
The magnitude of the second term in the integrand on the right-hand side of 
Eq.~(\ref{Mk}) is smaller than that of the first term 
by the factor $\sim \rho/L$. 
Except for this small correction, 
Eq.~(\ref{Mk}) agrees with the well-known expression of 
the magnetization vector.~\cite{H&M} 

For each particle species $a$, 
the same motion equations as shown in 
Eqs.~(\ref{uxieq})--(\ref{uvteq}) 
are derived from 
$\delta I_{DKF} / \delta x_{aE}^i = \delta I_{DKF} / \delta v_{\parallel aE}
= \delta I_{DKF} / \delta \mu_{aE} = \delta I_{DKF} / \delta \vartheta_{aE}
=  0$ 
and accordingly 
the same collisionless drift kinetic equation as Eq.~(\ref{DKE}) 
is obtained for the gyrophase-averaged distribution 
function $\overline{F}_a \equiv \oint F_a d\vartheta/(2\pi)$. 

The remaining governing equations of the system, namely, 
the quasineutrality condition and  Amp\`{e}re's law 
are derived from 
$\delta I_{DKF} / \delta \phi =  0$ 
and $\delta I_{DKF} / \delta A_i =  0$, respectively,  as 
\begin{equation}
\label{quasineutrality}
\sum_e e_a N_a
\equiv
\sum_a e_a 
\int d^3 v
\; F_a 
= 0
\end{equation}
and 
\begin{equation}
\label{ampere2}
\epsilon^{ijk}  \frac{\partial B_k }{\partial x^j} 
= \frac{4\pi}{c} J^i
, 
\end{equation}
where the $i$th contravariant component of the 
electric current vector density is defined by 
\begin{equation}
\label{Ji}
 J^i
\equiv 
\sum_a e_a N_a V_a^i 
\equiv 
\sum_a e_a 
\int d^3 v
 \; F_a  u_{ax}^i
+
c \,
\epsilon^{ijk}  \frac{\partial M_k }{\partial x^j} 
. 
\end{equation}
It is noted that 
the definitions of 
the density $N_a$ and the flow velocity $V_a^i$ 
which appear in Eqs.~(\ref{quasineutrality}) and (\ref{Ji}) 
are already shown in Eqs.~(\ref{N}) and (\ref{NV}), 
respectively. 

\subsection{The momentum conservation law}

We now consider the transformation of the spatial coordinates 
given by Eq.~(\ref{xprime}) again. 
Under the spatial coordinate transformation, 
the variables $(v_\parallel, \mu, \vartheta)$ are kept fixed 
as noted in Sec.~III.B.  
In the same way as in Eqs.~(\ref{dFDK3}) and (\ref{dupAg}), 
the spatial coordinate transformation causes
the variations in the distribution function $F_a$ and the functional forms  
of $(u_{a x}^i, u_{v_{a \parallel}}, u_{a \mu}, u_{a \vartheta})$ which 
are written as 
\begin{equation}
\overline{\delta} F_a 
=
- \frac{\partial}{\partial x^j}
( F_a \xi^j  )
, 
\end{equation}
and 
\begin{eqnarray}
& & 
\overline{\delta} u_{ax}^i
= 
u_{ax}^j \frac{\partial \xi^i}{\partial x^j} 
- \xi^j \frac{\partial u_{ax}^i}{\partial x^j}
, 
\hspace*{5mm}
\overline{\delta} u_{a v_\parallel}
=
- \xi^j \frac{\partial u_{a v_\parallel}}{\partial x^j}
, 
\nonumber \\ 
& & 
\overline{\delta} u_{a \mu}
=
- \xi^j \frac{\partial u_{a\mu}}{\partial x^j}
, 
\hspace*{5mm}
\overline{\delta} u_{a \vartheta}
= 
- \xi^j \frac{\partial u_{a\vartheta}}{\partial x^j}
.
\end{eqnarray}
 The variations in $\phi$, $A_i$, and $g_{ij}$ due to the spatial 
coordinate transformation are shown in 
Eqs.~(\ref{deltaphi})
(\ref{deltaA}), and (\ref{dEgg}), respectively.  

Using the expressions of these variations described above, 
we find that the variation $\overline{\delta} I_{DKF}$ in 
the action integral $I_{DKF}$ 
caused by the spatial coordinate transformation 
is written in the form, 
\begin{equation}
\label{dIDKF}
\overline{\delta} I_{DKF}
= 
\int_{t_1}^{t_2}  dt \int_V d^3 x 
\left[
\xi_j J_{DKF}^j
+
\frac{\partial }{\partial x^i} \left( \xi_j T_{DKF}^{ij}  \right)
\right]
. 
\end{equation}
Here,  $J_{DKF}^j$ is given by 
\begin{equation}
\label{Jj}
J_{DKF}^j
\equiv
\frac{\partial P_{tot}^j}{\partial t}
+
\nabla_i \Theta_{tot}^{ij}
, 
\end{equation}
  where $P_{tot}^j$ and $\Theta_{tot}^{ij}$ represent 
the total momentum vector density and
the total symmetric pressure tensor density defined by 
\begin{equation}
\label{Pc}
P_{tot}^j
\equiv 
g^{jk}
\sum_a \int d^3 v \, F_a \frac{\partial L_{GCa}}{\partial u_{ax}^k}
= 
\sum_a \int d^3 v \, F_a 
 m_a v_\parallel b^j 
, 
\end{equation}
and 
\begin{equation}
\label{tptf}
\Theta_{tot}^{ij}
 \equiv  
\Theta_p^{ij}
+ \Theta_f^{ij}
, 
\end{equation}
respectively. 
   It should be noted that, in Eq.~(\ref{Pc}), 
the vector potential part of the canonical momentum does not 
contribute to the total momentum because of the quasineutrality 
condition, Eq.~(\ref{quasineutrality}). 
   The first term on the right-hand side of Eq.~(\ref{tptf}) is 
the particle part of the pressure tensor density defined by
\begin{equation}
\label{Thetap}
\Theta_p^{ij}
 \equiv  
2 
\sum_a
\int d^3 v
\; F_a    
\frac{\partial L_{GCa}}{\partial g_{ij}}
= 
P_{\rm CGL}^{ij}
+ \pi_\land^{ij}
, 
\end{equation}
which consists of the CGL part,  
\begin{equation}
\label{CGL}
P_{\rm CGL}^{ij}
=
\sum_a \int d^3 v \, F_a
[ m_a v_\parallel^2 b^i b^j + \mu B ( g^{ij} - b^i b^j ) ]
, 
\end{equation}
and the non-CGL part,
\begin{equation}
\label{non-CGL}
\pi_\land^{ij}
\equiv 
\sum_a \int d^3 v \, F_a
m_a v_\parallel 
[ b^i ( u_{ax} )_\perp^j  + ( u_{ax} )_\perp^i b^j ]
.
\end{equation}
Equations~(\ref{Thetap}), (\ref{CGL}), and (\ref{non-CGL}) 
are just the species summation of 
Eqs.~(\ref{Pij}), (\ref{CGL0}), and (\ref{non-CGL0}), 
respectively. 
   The second term on the right-hand side of Eq.~(\ref{tptf})
is given by 
\begin{equation}
\label{Thetaf}
\Theta_f^{ij}
 \equiv  
2 \frac{\partial}{\partial g_{ij}}
\left(
- \frac{\sqrt{g}}{8 \pi} B^2
\right)
= 
\frac{\sqrt{g}}{4\pi}
\left( \frac{B^2}{2} g^{ij} - B^i B^j \right)
, 
\end{equation}
which represents the Maxwell stress tensor 
due to the magnetic field with the opposite sign. 
It is clear that $\Theta_{tot}^{ij}$, $\Theta_p^{ij}$, $\Theta_f^{ij}$, 
$P_{\rm CGL}^{ij}$, and $\pi_\land^{ij}$ are all symmetric 
with respect to the interchange of the superscripts $i$ and $j$.

The contravariant $(i,j)$-component $T_{DKF}^{ij}$ 
of the tensor density appearing on the 
left-hand side of Eq.~(\ref{dIDKF}) is written as  
\begin{equation}
\label{Tij}
T_{DKF}^{ij}
\equiv
\Pi_{tot}^{ij} - \Theta_{tot}^{ij}
-\nabla_k F_{tot}^{ijk} 
,
\end{equation}
where the total asymmetric canonical pressure tensor 
density $\Pi_{tot}^{ij}$ 
and the third-rank tensor density $F_{tot}^{ijk}$ are defined by
\begin{eqnarray}
\Pi_{tot}^{ij}
& \equiv & 
g^{jk}
\left[ 
\sum_a \int d^3 v \, F_a 
\left( u_{ax}^i 
\frac{\partial L_{Gca}}{\partial u_{ax}^k} 
- \nabla_k A_l
\frac{\partial L_{Gca}}{\partial (\partial A_l / \partial x^i)}
\right)
\right.
\nonumber \\ & & 
\left. \mbox{}
+ \nabla_k A_l
\frac{\partial}{\partial (\partial A_l / \partial x^i)}
\left(
\frac{\sqrt{g}}{8\pi} B^2
\right)
\right]
- g^{ij}
\frac{\sqrt{g}}{8\pi} B^2 
\nonumber \\
& = &
\sum_a \int d^3 v \, F_a 
u_{ax}^i 
\left(
m_a v_\parallel b^j 
+ \frac{e_a}{c} A^j
\right)
\nonumber \\ 
& & 
+ \frac{\epsilon^{ilm}}{4\pi} (B_m - 4\pi M_m) \nabla^j A_l
- g^{ij}
\frac{\sqrt{g}}{8\pi} B^2 
,
\end{eqnarray}
   and
\begin{eqnarray}
F_{tot}^{ijk} 
& \equiv & 
A^j
\left[ 
- 
\sum_a \int d^3 v \, F_a 
\frac{\partial L_{Gca}}{\partial (\partial A_k / \partial x^i)}
\right.
\nonumber \\ & & 
\left. \mbox{}
+ 
\frac{\partial}{\partial (\partial A_k/ \partial x^i)}
\left(
\frac{\sqrt{g}}{8\pi} B^2
\right)
\right]
\nonumber \\ 
& = & 
\frac{\epsilon^{ikm}}{4\pi} A^j  (B_m - 4\pi M_m)
,
\end{eqnarray}
respectively. 
We can immediately see that  $F_{tot}^{ijk}$ satisfies
\begin{equation}
F_{tot}^{ijk} = - F_{tot}^{kji}
, 
\end{equation}
from which we have 
\begin{equation}
\label{nnFz}
\nabla_i \nabla_k F_{tot}^{ijk} = 0
, 
\end{equation}
in the same way as in Eq.~(\ref{nablaik})

Since the action integral $I_{DKF}$ is invariant 
under the spatial coordinate transformation, 
$\overline{\delta} I_{DKF}$ written in Eq.~(\ref{dIDKF}) vanishes 
for any $\xi_j$. 
  Thus, the integrands at the interior and boundary 
points shown on the right-hand side of Eq.~(\ref{dIDKF}) 
should vanish 
separately so we obtain $J_{DKF}^j = 0$ and 
$T_{DKF}^{ij} = 0$. 
We find from Eqs.~(\ref{Jj}) and (\ref{Tij}) 
that $J_{DKF}^j = 0$ represents 
the total momentum conservation law, 
\begin{equation}
\label{momcons3}
\frac{\partial P_{tot}^j}{\partial t}
+
\nabla_i \Theta_{tot}^{ij}
= 0
, 
\end{equation}
and $T_{DKF}^{ij} = 0$ gives the relation of the 
total symmetric pressure tensor 
density $\Theta_{tot}^{ij}$ to the 
total asymmetric  canonical pressure tensor 
density $\Pi_{tot}^{ij}$, 
\begin{equation}
\label{thpic}
\Theta_{tot}^{ij} =
\Pi_{tot}^{ij} 
-\nabla_k F_{tot}^{ijk} 
.
\end{equation}
Combining Eqs.~(\ref{nnFz}) and (\ref{thpic}) shows  
\begin{equation}
\label{nablaT}
\nabla_i \Theta_{tot}^{ij} =
\nabla_i \Pi_{tot}^{ij} 
.
\end{equation}
We clearly see that 
the relations between the two types of the pressure tensors 
shown in Eqs.~(\ref{thpic}) and (\ref{nablaT}) take the same 
forms as those 
given by Eqs.~(\ref{thetapi}) and (\ref{nablaT0}) in Sec.~II.C, respectively.

It is noted that 
the momentum conservation law similar to 
Eq.~(\ref{momcons3}) was derived by 
Brizard and Tronci~\cite{B&T} for 
the guiding-center Vlasov-Maxwell system.  
In their model, the electromagnetic fields are 
determined by the full Maxwell equations 
including the Maxwell displacement current 
so that their system contains 
such rapid phenomena as the electromagnetic waves 
with the speed of light and the Maxwell stress 
due to the electric field.
They also derived the symmetric pressure tensor including 
the same particle part as given by Eqs.~(\ref{Thetap})--(\ref{non-CGL}) 
although they modified the magnetization term in the canonical momentum 
conservation law to transform 
the asymmetric pressure tensor to the symmetric one.  
Thus, their derivation is different from our direct derivation of 
the symmetric pressure tensor by taking the variation with respect to 
the metric tensor.

It is instructive here to consider momentum conservation law, 
Eq.~(\ref{momcons3}), in the the equilibrium limit where 
the distribution functions are assumed to take the local Maxwellian form, 
$F_a = N_a ( m_a / 2 \pi T_a)^{3/2} \exp 
[- (\frac{1}{2} m_a v_\parallel^2 + \mu B ) / T_a ]$ 
(note that, precisely speaking, this local Maxwellian distribution function is not the 
exact stationary solution but the zeroth-order one 
of the drift kinetic equation in the gyroradius ordering and that 
the deviation from the local Maxwellian appears in the first-order 
solution). 
Then, it is found from Eqs.~(\ref{Thetap})--(\ref{non-CGL}) that 
$\Theta_p^{ij} = P g^{ij}$ where $P \equiv \sum_a N_a T_a$. 
We now use the conventional vector notation 
to rewrite Eq.~(\ref{momcons3}) 
in the equilibrium state ($\partial/\partial t = 0$) as
\begin{equation}
\nabla P - \frac{1}{4\pi} (\nabla \times {\bf B} ) \times {\bf B}
= 0
,
\end{equation}
where Eqs.~(\ref{tptf}) and (\ref{Thetaf}) are used. 
In addition, Amp\`{e}re's law in Eq.~(\ref{ampere2}) is used to obtain 
the familiar force balance equation in 
the magnetohydrodynamics (MHD) equilibrium,  
\begin{equation}
\nabla P = \frac{1}{c} {\bf J} \times {\bf B},
\end{equation}
where the current density is given by Eq.~(\ref{Ji}) as 
\begin{equation}
{\bf J}
=
\sum_a e_a 
\int d^3 v
 \; F_a  {\bf u}_{ax}
+
c \nabla \times  {\bf M} 
. 
\end{equation}
This formula is called the magnetization law~\cite{H&M}; 
the current is represented by the sum of 
the flow of guiding centers and the curl of the magnetization ${\bf M}$ 
[Eq.~(\ref{Mk})], which are given by the first and 
second terms on the right-hand side, respectively. 
As shown in Ref.~\cite{H&M}, 
it is found from using the local Maxwellian distribution functions 
that the sum of 
the perpendicular components of the first and second terms on the 
right-hand side of the above magnetization law gives 
the diamagnetic current, 
$(c/B^2) ({\bf B} \times \nabla P)$. 
Recall that the perpendicular component 
$({\bf u}_{ax})_\perp$ of the guiding center velocity 
and the magnetization ${\bf M}$ are both produced from gyrations of 
particles around magnetic field lines. 
Even though finite gyroradius effects are not described by  
the guiding center distribution functions $F_a$ alone, 
such effects are partly included in $({\bf u}_{ax})_\perp$ and 
${\bf M}$ which help express the current properly 
and recover the familiar force balance equation in the MHD equilibrium 
as shown above.

\section{EFFECTS OF COLLISIONS}

We here investigate how collisions influence the momentum conservation laws
and the momentum balance equation shown in Secs.~II.C, III.C, and IV.B
when the collision term is added into the right-hand sides of 
the Vlasov and drift kinetic equations. 
   Effects of the collision term added into the right-hand side of 
Eq.~(\ref{Vlasov}) were already studied   
in Ref.~\cite{Sugama2015}
where it was shown how to evaluate the correction of 
the energy and momentum conservation laws due to the collision 
and other source terms.  
   According to the prescription given in Ref.~\cite{Sugama2015}, 
the modified conservation laws are obtained from the original ones 
with the time derivative of the distribution function being replaced as 
\begin{equation}
\label{replace}
\frac{\partial F_a}{\partial t}
\hspace*{3mm}
\rightarrow
\hspace*{3mm}
\frac{\partial F_a}{\partial t} - {\cal K}_a
, 
\end{equation}
where ${\cal K}_a$ is the term added into the 
the right-hand side of the kinetic equation to 
represent the rate of change in the distribution function 
$F_a$ due to Coulomb collisions 
and it may also include other parts representing 
external particle, momentum, and/or energy sources if any.

When ${\cal K}_a$ is added into the right-hand side of Eq.~(\ref{Vlasov}), 
Eq.~(\ref{replace}) is applied to the momentum conservation law in Eq.~(\ref{momcons}), where 
the term $\partial P_c^j/\partial t$ contains $\sum_a \partial F_a/\partial t$ 
as seen from Eq.~(\ref{Pcj}). 
Then, we find that the resulting momentum balance equation is given by 
Eq.~(\ref{momcons}) with making the replacement, 
\begin{equation}
\label{replace2}
\frac{\partial P_c^j}{\partial t}
\hspace*{3mm}
\rightarrow
\hspace*{3mm}
\frac{\partial P_c^j}{\partial t} 
- 
\sum_a \int d^3 v \, {\cal K}_a
\left( m_a v^j + \frac{e_a}{c} A^j \right)
. 
\end{equation}
In the case where ${\cal K}_a$ is given by the Coulomb collision operator 
(such as the Landau operator)
which satisfies the conservation laws of the particles' number 
($\int d^3 v \, {\cal K}_a = 0$) and 
the momentum  ($\sum_a \int d^3 v \, {\cal K}_a m_a v^j = 0$), 
the velocity space integral vanishes in Eq.~(\ref{replace2}) and   
we have the momentum conservation law in the same form 
as that for the case of ${\cal K}_a = 0$. 
Also, it is noted in Ref.~\cite{Sugama2015} that, even if 
${\cal K}_a$ contains some external source parts other than the collision term, 
the charge conservation law requires the condition 
$\sum_a e_a \int d^3 v \,  {\cal K}_a  = 0$ which implies 
the correction term proportional to $A^j$ vanishes in Eq.~(\ref{replace2}).  

Next, let us consider the case where
${\cal K}$ is added into the right-hand side of the drift kinetic equation, 
Eq.~(\ref{DKE}), for a given particle species, 
in which the subscript representing the particle species is omitted. 
Here,  ${\cal K}$ is regarded as gyrophase-averaged. 
Applying Eq.~(\ref{replace}) to this case, we find the momentum balance 
equations is derived from Eq.~(\ref{mombal}) with the following replacement, 
\begin{equation}
\label{replace3}
\frac{\partial}{\partial t} 
\left(  m N V_{g \parallel} b^j \right)
\hspace*{3mm}
\rightarrow
\hspace*{3mm}
\frac{\partial}{\partial t} 
\left(  m N V_{g \parallel} b^j \right)
- 
\int d^3 v \, {\cal K} 
m v_\parallel b^j
.
\end{equation}
The parallel component of this derived momentum balance equation 
agrees with Eq.~(18) in Ref.~\cite{Sugama2016} 
where its perpendicular components are not derived. 
We see from Eq.~(\ref{replace3}) that 
the effect of ${\cal K}$ on the momentum balance equation 
for the single particle species is written as  
$\int d^3 v \, {\cal K} m v_\parallel b^j$. 
When ${\cal K}$ is given by the Coulomb collision operator, 
$\int d^3 v \, {\cal K} m v_\parallel b^j$ 
represents the collisional transfer of 
the parallel momentum from 
the other particle species to the given species. 

Since the momentum balance equation obtained by 
substituting Eq.~(\ref{replace3}) into Eq.~(\ref{mombal}) 
is always valid for the distribution function which is the solution of the
drift kinetic equation including ${\cal K}$, 
it is also valid for each particle species 
even when the quasineutrality condition and Amp\`{e}re's law  
are additionally imposed for the self-consistent fields as in Sec.~IV. 
Furthermore, we can use Eq.~(\ref{replace}) in Eq.~(\ref{momcons3}) to 
see how the total momentum conservation law for the drift kinetic system in 
the self-consistent fields are modified by adding ${\cal K}_a$ into the 
drift kinetic equation for the particle species $a$. 
The resultant momentum balance equation is given from Eq.~(\ref{momcons3}) 
by putting 
\begin{equation}
\label{replace4}
\frac{\partial P_{tot}^j}{\partial t}
\hspace*{3mm}
\rightarrow
\hspace*{3mm}
\frac{\partial P_{tot}^j}{\partial t} 
- 
\sum_a \int d^3 v \, {\cal K}_a
m_a v_\parallel b^j 
. 
\end{equation}
   This corresponds to the species summation of Eq.(\ref{replace3}).  
When ${\cal K}_a$ represents the Coulomb 
collision operator in the zero-gyroradius limit, 
it satisfies $\sum_a \int d^3 v \, {\cal K}_a m_a v_\parallel = 0$ and  
the momentum conservation law 
takes the same form as in Eq.~(\ref{momcons3}). 
Note that the momentum conservation in Coulomb collisions is satisfied locally 
at the colliding particles' position which differs from the guiding-center position. 
Therefore, 
if the finite gyroradius effect is taken into account, 
$\sum_a \int d^3 v \, {\cal K}_a m_a v_\parallel$ does not generally vanish 
for the gyrophase-averaged collision 
operator ${\cal K}_a$ at the fixed guiding-center position, 
which includes the classical transport processes.~\cite{Sugama2015}

\section{CONCLUSIONS}

In this work, Eulerian variational formulations 
for kinetic plasma systems are presented.
As examples, 
the Vlasov-Poisson-Amp\`{e}re system and the drift kinetic systems are 
investigated. 
For the drift kinetic system, the additional case is also considered 
in which 
the quasineutrality condition and Amp\`{e}re's law are 
included as supplementary governing equations to describe 
the self-consistent fields. 

For all cases treated here, general spatial coordinates 
are used to represent the action integrals and 
the governing equations which take the forms being 
invariant under an arbitrary  (time-independent) transformation 
of spatial coordinates. 
   Furthermore, the invariance of the action integral under the spatial 
coordinate transformation is made use of to derive
the momentum conservation laws and/or the momentum balance
in which the functional derivatives of the Lagrangians 
with respect to the metric tensor components 
yield the proper symmetric pressure tensors 
more directly than conventional techniques using 
translational and rotational symmetries or taking the moments of 
the kinetic equations. 
 
  It is also clarified how the momentum balances are influenced 
by adding the collision and/or external source terms into the 
kinetic equations.  
Since the invariance under the spatial coordinate transformations 
is valid independently whether the system has symmetric geometry or not, 
the present formulation can be applied to 
kinetic studies of 
plasmas confined in general magnetic configurations including nonaxisymmetric systems such as stellarators and heliotrons.~\cite{Wakatani} 
  For example, 
the momentum balance equation derived here for the drift kinetic system 
are considered useful for verifications of accuracy of numerical simulations 
using Littlejohn's guiding center equations to study 
neoclassical transport processes in various magnetic geometries.  
  The extension of the present study to the gyrokinetic system is 
now in progress and the results will be reported elsewhere.

\begin{acknowledgments}
This work is supported in part by JSPS Grants-in-Aid for Scientific Research Grant Number 16K06941 and in part by the NIFS Collaborative Research Program NIFS16KNTT035. 
\end{acknowledgments}

\appendix

\section{MOMENTUM BALANCE IN THE VLASOV-POISSON SYSTEM}

We here consider the Vlasov-Poisson system, in which 
the electrostatic approximation holds; 
the magnetic filed is externally given as a time-independent one, 
${\bf B}_0  ({\bf x}) = \nabla \times {\bf A}_0  ({\bf x})$, 
and the electric field is written in 
terms of the electrostatic potential $\phi ({\bf x}, t)$ as 
${\bf E} ({\bf x}, t) = - \nabla \phi ({\bf x}, t)$. 
   The action integral 
$I_{VP}$ to describe the Vlasov-Poisson system is given by  
\begin{equation}
\label{IVP}
I_{VP}  \equiv  \int_{t_1}^{t_2} dt \; L_{VP} 
\equiv \int_{t_1}^{t_2}  dt \int_V d^3 x\; {\cal L}_{VP} 
,
\end{equation}
where the Lagrangian density ${\cal L}_{VP}$ is written as 
\begin{eqnarray}
\label{LVPdens} 
{\cal L}_{VP} 
& \equiv & 
\sum_a 
\int d^3 v \; F_a (x^i, v^i, t) L_a  +  {\cal L}_{VPf} 
.
\end{eqnarray}
Here, the single-particle Lagrangian $L_a$ for species $a$ is 
defined by Eq.~(\ref{spL}) where the covariant components $A_j (x^n, t)$ 
of the vector potential are replaced with the time-independent ones 
$A_{0j} (x^n)$.  
The field Lagrangian density ${\cal L}_{VPf}$ is written in the 
general spatial coordinates  $(x^i)_{i =1,2,3}$ 
as 
\begin{equation}
\label{LVPfdens} 
{\cal L}_{VPf} 
\equiv  
\frac{\sqrt{g(x^n)}}{8\pi} g^{ij}(x^n)
\frac{\partial \phi (x^n, t)}{\partial x^i}
\frac{\partial \phi (x^n, t)}{\partial x^j}
.
\end{equation}
%

  In the same way as in Sec.~II.A, 
we now consider the virtual displacement of the particle's trajectory, 
for which the variations in the particle's position and velocity
are represented in the Eulerian picture as 
$\delta x_{a E}^i$ and $\delta v_{a E}^i$, respectively 
[see Eq.~(\ref{deltaxvE})]. 
The electrostatic potential field $\phi$ is also virtually varied by 
$\delta \phi$.  
However, since the vector potential $A_{0j}$ is fixed, 
its virtual variation $\delta A_{0j}$ does not appear.  
Then, the variation in the action integral $I_{VP}$ is given by 
\begin{eqnarray}
\label{deltaIVP}
\delta I_{VP} 
& = & 
\sum_a
\int_{t_1}^{t_2} dt 
\int d^3 x \int d^3 v \; F_a 
\left[
\delta x_{aE}^i \left\{ 
\left( \frac{\partial L_a}{\partial x^i} \right)_{u_{ax}}
\right. \right. 
\nonumber \\
& & \left. \left. \mbox{}
-  \left(  \frac{d}{d t } \right)_a
\left( \frac{\partial L_a}{\partial u_{ax}^i} \right)
\right\}
+ \delta v_{aE}^i 
\left(\frac{\partial L_a}{\partial v^i}\right)_{u_{ax}}
\right]
\nonumber \\
& & \mbox{}
\hspace*{-4mm}
+ \int_{t_1}^{t_2} dt \int d^3 x 
\;
\delta \phi \left(
- \sum_a e_a \int d^3 v \; F_a 
- \frac{\sqrt{g}}{4\pi} \Delta \phi 
\right)
\nonumber \\
& & 
 \mbox{}
\hspace*{-4mm}
+ \delta I_{VPb} 
, 
\end{eqnarray}
where  
$\delta I_{VPb}$ represents the part which is determined from the values 
of $\delta x_{aE}^i$ and $\delta \phi$ on the boundaries of the integral region. 
  In deriving Eq.~(\ref{deltaIVP}), 
Eqs.~(\ref{duvax}) and (\ref{dFa}) are used. 
Imposing the condition that $\delta I_{VP}=0$  for arbitrary variations 
$\delta x_{aE}^i$, $\delta v_{aE}^i$ and $\delta \phi$ which vanish on the 
boundaries, the same equations as those in 
Eqs.~(\ref{Vlasov}) and (\ref{poisson}) are 
obtained in the same manner as shown in Sec.~II.A. 
Recalling that, in the present case, 
$E_i = - \partial \phi / \partial x^i$ because of 
$\partial A_{0i}/\partial t = 0$, 
we confirm the fact that Eqs.~(\ref{Vlasov}) and (\ref{poisson}) resulting from 
$\delta I_{VP}=0$ 
form the governing equations of the Vlasov-Poisson system. 

   To derive the momentum balance in the Vlasov-Poisson system, 
we next consider the infinitesimal spatial coordinate transformation as shown in 
Eq.~(\ref{xprime}) of Sec.~II.B. 
In the same way as in Sec.~II.B, 
the variations in 
$v^i$, $\phi$, $F_a$, $u_{ax}^i$, and $u_{av}^i$ due to the spatial coordinate 
transformation are denoted by $\overline{\delta} v^i $, $\overline{\delta} \phi$, 
$\overline{\delta} F_a$, $\overline{\delta} u_{ax}^i$, and 
$\overline{\delta} u_{av}^i$, respectively, which 
are defined by 
Eqs.~(\ref{ovldv}), (\ref{deltaphi}), (\ref{dFxieta}), and (\ref{duxieta}). 
We should note that the spatial coordinate transformation also 
causes the variations in the metric tensor components
 [see Eq.~(\ref{dEgg})] 
as well as the variation $\overline{\delta} A_{0i}$ in the functional form of the contravariant component 
$A_{0i}$ of the externally given vector potential 
where $\overline{\delta} A_{0i}$ is written in the same 
form as in Eq.~(\ref{deltaA}),  
\begin{equation}
\label{deltaA0}
\overline{\delta} A_{0i} 
= 
 - \xi^j  \frac{\partial A_{0i}}{\partial x^j} 
- \frac{\partial \xi^j}{\partial x^i}
A_{0j}
\equiv 
-  L_\xi  A_{0i}
. 
\end{equation}
  This is contrast to the case 
that $\delta A_{0i}$ does not appear 
in considering the virtual variations to derive
the governing equations of the Vlasov-Poisson system 
from $\delta I_{VP} = 0$. 
     Using Eqs.~(\ref{deltaphi}),  
(\ref{dEgg}), (\ref{dFxieta}), (\ref{duxieta}), and (\ref{deltaA0}), 
it is found that the variation $\overline{\delta} I_{VP}$ in the action integral $I_{VP}$ 
due to the spatial coordinate transformation is written as  
\begin{equation}
\label{dIVP}
\overline{\delta} I_{VP}
=
\int_{t_1}^{t_2} dt 
\int_V d^3 x 
\left[ 
\xi_j 
J_{VP}^j
+ \frac{\partial }{\partial x^i}
(\xi_j T_{VP}^{ij} ) \right]
,
\end{equation}
where 
$J_{VP}^j$ and $T_{VP}^{ij}$ are given by 
\begin{eqnarray}
\label{JVP}
J_{VP}^j
& \equiv & 
\frac{\partial P_c^j}{\partial t} 
+ 
A_0^j 
\frac{\partial}{\partial x^k} 
\left( 
\sum_a \int d^3 v \, F_a 
\frac{\partial L_a}{\partial A_{0k}}
\right)
+ \nabla_i \Pi^{ij}
- F_L^j
\nonumber \\
& = & 
\frac{\partial P^j}{\partial t} 
+ \nabla_i \Pi^{ij}
- F_L^j
,
\end{eqnarray}
and
\begin{equation}
\label{TVP}
T_{VP}^{ij}
\equiv 
 \Pi_c^{ij} - \Pi^{ij} 
- A_0^j \sum_a \int d^3 v \, F_a 
\frac{\partial L_a}{\partial A_{0i}}
,
\end{equation}
respectively. 
    The conditions,
$\delta I / \delta x_{aE}^i = \delta I / \delta v_{aE}^i = 0$ and
$\delta I / \delta \phi = 0$, from which the Vlasov kinetic equation and 
Poisson's equation are derived, are also used 
in deriving Eq.~(\ref{dIVP}). 
  In Eq.~(\ref{JVP}), 
$P^j$ and $P_c^j$ are  
the kinetic and canonical momentum densities which are defined by 
\begin{equation}
\label{PjVP}
P^j 
 \equiv 
\sum_a \int d^3 v \; F_a 
m_a v^j 
,
\end{equation}
and 
\begin{equation}
\label{PcjVP}
P_c^j 
 \equiv 
g^{jk} 
\sum_a \int d^3 v \, F_a 
\frac{\partial L_a}{\partial u_{ax}^k}
= 
\sum_a \int d^3 v \; F_a 
\left( m_a v^j + \frac{e_a}{c} A_0^j \right)
,
\end{equation}
respectively, and  
$F_L^j$ represents the Lorentz force given by 
\begin{eqnarray}
\label{FL}
F_L^j
& \equiv & 
g^{ij} \left( \frac{\partial A_{0k}}{\partial x^i}
- \frac{\partial A_{0i}}{\partial x^k} \right)
\sum_a \int d^3 v \, F_a 
\frac{\partial L_a}{\partial A_{0k}}
\nonumber \\
& = & 
\frac{\epsilon^{jkl}}{\sqrt{g}} 
\left( \sum_a \frac{e_a}{c}
\int d^3 v \; F_a v_k \right) 
B_{0l} 
,
\end{eqnarray}
where 
$B_{0i} \equiv g_{ij} B_0^j$,  
$B_0^i \equiv  ( \epsilon^{ijk} / \sqrt{g} )
(\partial A_{0k} / \partial x^j )$, 
and 
$\partial L_a / \partial A_{0k} 
= (e_a / c) v^k$ are used. 
Using the continuity equation derived from the Vlasov kinetic equation, 
we can confirm that the right-hand side of the first line in Eq.~(\ref{JVP}) 
equals the last line.  
   In addition, Eqs.~(\ref{JVP}) and (\ref{TVP}) contain 
the symmetric pressure tensor $\Pi^{ij}$ and  
the canonical pressure tensor $\Pi_c^{ij}$ 
which are defined  by 
\begin{eqnarray}
\label{PijVP}
\Pi^{ij}
 & \equiv & 
2 
\left( 
\sum_a \int d^3 v \; F_a 
\frac{\partial L_a}{\partial g_{ij}} 
+
\frac{\partial {\cal L}_{VPf}}{\partial g_{ij}}
\right) 
\nonumber \\
& = & 
\sum_a \int d^3 v \; F_a m_a v^i v^j
+ \frac{\sqrt{g}}{4\pi}
\left ( 
\frac{g^{ij}}{2} E_L^k E_{Lk} - E_L^i E_L^j 
\right) 
,
\nonumber \\ & & 
\end{eqnarray}
and 
\begin{eqnarray}
\label{PcijVP}
\Pi_c^{ij}
&  \equiv & 
g^{jk} 
\left ( \sum_a \int d^3 v \; F_a v^i \frac{\partial L_a}{\partial u_{ax}^k}
- \frac{\partial \phi}{\partial x^k} 
\frac{\partial {\cal L}_{VPf}}{\partial (\partial \phi / \partial x^i )}
\right) 
\nonumber \\
& & \mbox{} 
+ \frac{\sqrt{g}}{8\pi} g^{ij} g^{kl} 
\frac{\partial \phi}{\partial x^k} \frac{\partial \phi}{\partial x^l} 
\nonumber \\
& = & 
\sum_a \int d^3 v \; F_a v^i 
\left( m_a v^j + \frac{e_a}{c} A_0^j \right)
\nonumber \\
& & \mbox{} 
+ \frac{\sqrt{g}}{4\pi}
\left( 
\frac{g^{ij}}{2} E_L^k E_{Lk} - E_L^i E_L^j 
\right) 
,
\end{eqnarray}
respectively, 
where 
$E_L^i \equiv g^{ij} E_{Lj}$ and 
$E_{Li} \equiv - \partial \phi / \partial x^i$ are used. 

   Because of the invariance of the action integral $I_{VP}$ under the 
general spatial coordinate transformation, 
$\overline{\delta} I_{VP}$ vanishes for any $\xi_j$ , and accordingly, 
we have $J_{VP}^j = 0$ and $T_{VP}^{ij} = 0$ 
from Eq.~(\ref{dIVP}). 
   The momentum balance in the Vlasov-Poisson system is obtained from 
$J_{VP}^j = 0$ as 
\begin{equation}
\label{mbVP}
\frac{\partial P^j}{\partial t} 
+ \nabla_i \Pi^{ij}
=
 F_L^j
,
\end{equation}
which agrees with that shown by Qin {\it et al}.~\cite{Qin2} 
Another condition $T_{VP}^{ij} = 0$ gives the relation between 
the symmetric pressure tensor $\Pi^{ij}$ and the canonical pressure 
tensor $\Pi_c^{ij}$. 
The validity of $T_{VP}^{ij} = 0$ is also easily verified from  
Eqs.~(\ref{TVP}), (\ref{PijVP}), (\ref{PcijVP}), and 
$\partial L_a / \partial A_{0i} 
= (e_a / c) v^i$. 

\section{ENERGY CONSERVATION IN THE VLASOV-POISSON SYSTEM}
In this Appendix, 
we  consider the energy balance in the Vlasov-Poisson system. 
The energy conservation laws for 
the Vlasov-Poisson-Amp\`{e}re system and 
the Boltzmann-Poisson-Amp\`{e}re system 
are shown in Refs.~\cite{Sugama2013} and \cite{Sugama2015}, 
respectively.   
In contrast to the case in Appendix~A where the momentum balance 
in the Vlasov-Poisson system is derived, 
we do not need to use the general spatial coordinate system here.  
So we now use only the Cartesian coordinate system and represent 
three-dimensional vectors in terms of boldface letters. 
Either a Lagrangian or an Eulerian variational formulation can be 
used for the derivation of the energy balance 
although we here follow the Eulerian formulation to 
treat the variation of the action integral under translation in time. 
  The infinitesimal time translation is represented by 
transforming the time coordinate as
\begin{equation}
\label{time_translation}
t' = t + \epsilon
,
\end{equation}
where $\epsilon$ is an infinitesimal constant. 
  The time translation causes the variations 
$\delta_t I_{VP}$ in the action integral $I_{VP}$, 
where $I_{VP}$ is defined in Eq.~(\ref{IVP}) and 
$\delta_t I_{VP}$ is written as 
\begin{eqnarray}
\label{AdIVP}
\delta_t I_{VP} 
& = & 
\int_{t_1}^{t_2} \int_V d^3 x 
\left[ 
\epsilon \frac{\partial {\cal L}_{VP} }{\partial t}
+ \sum_a \int d^3 v 
\right. 
\nonumber \\ 
& & \mbox{} 
\hspace*{-5mm} 
\times 
\left\{
\delta_t F_a \cdot L_a 
+
 F_a \left(
\frac{\partial L_a }{\partial {\bf u}_{ax}} 
\cdot  \delta_t {\bf u}_{ax}
+ \frac{\partial L_a}{\partial  {\bf u}_{av}}  
\cdot   \delta_t {\bf u}_{av}
\right. \right. 
\nonumber \\ 
& & 
\left. \left. \left. 
+  \frac{\partial L_a}{\partial \phi}
\delta_t \phi
\right)
\right\}
+  \frac{\partial {\cal L}_{VPf} }{\partial \nabla\phi}
\cdot
\nabla \delta_t \phi
\right]
. 
\end{eqnarray}
In this Appendix, 
we use $\delta_t \cdots$ to represent the 
variations associated with the time translation. 
The variations in 
${\bf u}_{ax} \equiv (u_{ax}^i)_{i=1,2,3}$,  
${\bf u}_{av} \equiv (u_{av}^i)_{i=1,2,3}$,  
$\phi$, 
and $F_a$ due to the time translation are written as 
\begin{equation}
\label{Adua}
\delta_t {\bf u}_{ax} 
= - \epsilon \frac{\partial {\bf u}_{ax}}{\partial t}
,  \hspace*{3mm} 
\delta_t {\bf u}_{av} 
= - \epsilon \frac{\partial {\bf u}_{av}}{\partial t}
,  \hspace*{3mm} 
\delta_t \phi 
= - \epsilon \frac{\partial \phi}{\partial t}
\end{equation}
and 
\begin{equation}
\label{AdFa}
\delta_t F 
=  - \epsilon \frac{\partial F}{\partial t}
=
\epsilon 
\left[
\frac{\partial}{\partial {\bf x}} \cdot
( F {\bf u}_{ax}  )
+ \frac{\partial}{\partial {\bf v}} \cdot
( F {\bf u}_{av}  )
\right]
, 
\end{equation}
respectively, 
where Eq.~(\ref{Vlasov0}) is used.  
Then, substituting Eqs.~(\ref{Adua}) and (\ref{AdFa}) into 
Eq.~(\ref{AdIVP}) and using 
$\delta I_{VP}/\delta {\bf x}_E = \delta I_{VP}/\delta {\bf v}_E =0$ and 
$\delta I_{VP}/\delta \phi =0$, 
we obtain  
\begin{equation}
\label{AdIVP2}
\delta_t I_{VP} 
=
- \epsilon \int_{t_1}^{t_2} \int_V d^3 x 
\left( 
\frac{\partial {\cal E}_{VPc}}{\partial t} 
+ \frac{\partial }{\partial {\bf x} } 
\cdot 
{\bf Q}_{VPc}
\right)
, 
\end{equation}
where  the canonical energy density 
${\cal E}_{VPc}$ and the canonical energy flux 
${\bf Q}_{VPc}$ are defined by 
\begin{equation}
\label{EVPc}
{\cal E}_{VPc} 
\equiv 
\sum_a \int d^3 v \, F_a 
\left( 
\frac{1}{2} m v^2 + e_a \phi 
\right)
- \frac{| {\bf E}_L |^2}{8\pi}
, 
\end{equation}
and 
\begin{equation}
\label{QVPC}
{\bf Q}_{VPc}
 \equiv  
\sum_a \int d^3 v \, F_a 
\left( 
\frac{1}{2} m v^2 + e_a \phi  
\right) {\bf v}
- \frac{1}{4\pi}
\frac{\partial \phi}{\partial t}
{\bf E}_L
,
\end{equation}
respectively. 
Here, the electrostatic electric field is 
represented by 
${\bf E}_L \equiv - \nabla \phi$. 

Since the Lagrangian density ${\cal L}_{VP}$ defined in 
Eq.~(\ref{LVPdens}) for the 
Vlasov-Poisson system 
depends on time $t$ 
only through the functions ${\bf u}_{ax}$, $F_a$, and $\phi$ 
which are all determined by the variational principle (see Appendix~A),  
the action integral $I_{VP}$ given in Eq.~(\ref{IVP}) is invariant 
under the time translation. 
   Therefore, $\delta_t I_{VP}$ 
vanishes for an arbitrarily chosen  
 integral region $[t_1, t_2] \times V$ 
and accordingly, the integrand in Eq.~(\ref{AdIVP2}) also vanishes. 
 Thus, we obtain the local energy conservation law written as 
\begin{equation}
\label{AECVP}
\frac{\partial {\cal E}_{VPc}}{\partial t} 
+ \frac{\partial }{\partial {\bf x} } 
\cdot 
{\bf Q}_{VPc}
= 
\frac{\partial {\cal E}_{VP}}{\partial t} 
+ \frac{\partial }{\partial {\bf x} } 
\cdot 
{\bf Q}_{VP}
= 0
,
\end{equation}
where the energy density 
${\cal E}_{VP}$ and the energy flux 
${\bf Q}_{VP}$ are defined by 
\begin{equation}
\label{EVP}
{\cal E}_{VP} 
\equiv 
\sum_a \int d^3 v \, F_a 
\frac{1}{2} m v^2
+ \frac{| {\bf E}_L |^2}{8\pi}
, 
\end{equation}
and 
\begin{equation}
\label{QVP}
{\bf Q}_{VP}
 \equiv  
\sum_a \int d^3 v \, F_a 
\left( 
\frac{1}{2} m v^2 + e_a \phi  
\right) {\bf v}
+ \frac{\phi}{4\pi}
 \frac{\partial {\bf E}_L}{\partial t} 
, 
\end{equation}
respectively. 
Poisson's equation shown in Eq.~(\ref{poisson}) is also used
for deriving Eq.~(\ref{AECVP}). 
The local energy conservation law shown in 
Eq.~(\ref{AECVP}) agrees with that obtained by 
Qin {\it et al}.~\cite{Qin2}

\section{ENERGY BALANCE IN THE DRIFT KINETIC SYSTEM}

The energy balance in the drift kinetic system is considered 
in this Appendix.  
The infinitesimal time translation shown in  
Eq.~(\ref{time_translation}) causes the variations 
$\delta_t I_{DK}$ in the action integral $I_{DK}$, 
where $I_{DK}$ is defined in Eq.~(\ref{IDK}).  
and $\delta_t I_{DK}$ is written as
%
\begin{eqnarray}
\label{AdIDK}
\delta_t I_{DK} 
& = & 
\int_{t_1}^{t_2} \int_V d^3 x 
\left[ 
\epsilon \frac{\partial {\cal L}_{DK} }{\partial t}
+ \int d^3 v \left\{
 F \left(
\frac{\partial L_{GC} }{\partial {\bf u}_x} 
\cdot  \delta_t {\bf u}_x
\right.  \right. \right. 
\nonumber \\ & & \mbox{}
\left.  \left. \left. 
+ \frac{\partial L_{GC}}{\partial  u_\vartheta}  
 \delta_t u_\vartheta
\right)
+  \delta_t F \cdot L_{GC} 
\right\}
\right]
. 
\end{eqnarray}
%
In the same way as in Appendix~B, 
we here use $\delta_t \cdots$ to denote the 
variations associated with the time translation. 
We also use the Cartesian coordinate system  
and represent three-dimensional vectors in terms of boldface letters. 
The variations in 
${\bf u}_x \equiv (u_x^i)_{i=1,2,3}$,  
$u_{v_\parallel}$, $u_\mu$, $u_\vartheta$, 
and $F$ due to the time translation are written as 
\begin{eqnarray}
\label{Adu}
\delta_t {\bf u}_x 
= - \epsilon \frac{\partial {\bf u}_x}{\partial t}
, & \hspace*{3mm} & 
\delta_t u_{v_\parallel} 
= - \epsilon \frac{\partial u_{v_\parallel}}{\partial t}
, 
\nonumber \\  
\delta_t u_\mu 
= - \epsilon \frac{\partial u_\mu}{\partial t}
, & \hspace*{3mm} & 
\delta_t u_\mu 
= - \epsilon \frac{\partial u_\mu}{\partial t}
, 
\end{eqnarray}
and 
\begin{eqnarray}
\label{AdF}
\delta_t F 
=  - \epsilon \frac{\partial F}{\partial t}
& = & 
\epsilon 
\left[
\frac{\partial}{\partial {\bf x}} \cdot
( F {\bf u}_x  )
+ \frac{\partial}{\partial v_\parallel}
( F u_{v_\parallel}  )
\right. 
\nonumber \\ & & 
\left. \mbox{}
+ \frac{\partial}{\partial \mu}
( F u_\mu )
+ \frac{\partial}{\partial \vartheta}
( F u_\vartheta  )
\right]
, 
\end{eqnarray}
respectively, 
where Eq.~(\ref{DKE0}) is used.  
Substituting Eqs.~(\ref{Adu}) and (\ref{AdF}) into 
Eq.~(\ref{AdIDK}) and using 
$\delta I_{DK}/\delta {\bf x}_E =0$ [Eq.~(\ref{dpidt})], 
$\delta I_{DK}/\delta v_{\parallel E} =0$ [Eq.~(\ref{dLdv})], 
$\delta I_{DK}/\delta \mu_E =0$ [Eq.~(\ref{dIdmu})], 
and $\delta I_{DK}/\delta \vartheta_E =0$ [Eq.~(\ref{dIdth})], 
we can rewrite $\delta_t I_{DK}$ as  
\begin{eqnarray}
\label{AdIDK2}
\delta_t I_{DK} 
& = & 
- \epsilon \int_{t_1}^{t_2} \int_V d^3 x 
\left[ 
\frac{\partial }{\partial t} 
\left(
\int d^3 v \, F {\cal E} 
\right)
\right. 
\nonumber \\ & & 
\left. \mbox{}
+ \frac{\partial }{\partial {\bf x} } 
\cdot 
\left(
\int d^3 v \, 
F {\cal E} {\bf u}_x
 \right)
\right]
, 
\end{eqnarray}
where  
the guiding center velocity ${\bf u}_x$ is given by  
Eq.~(\ref{uxieq}) 
and ${\cal E}$ represents 
the energy 
of the single particle 
(or the guiding-center Hamiltonian $H_{GC}$) 
defined by  
\begin{eqnarray}
{\cal E} 
& \equiv & 
H_{GC} 
\equiv
\frac{\partial L_{GC} }{\partial {\bf u}_x} 
\cdot {\bf u}_x
+ \frac{\partial L_{GC}}{\partial  u_\vartheta}  
 u_\vartheta
-  L_{GC}
\nonumber \\
& = & 
\frac{1}{2} m v_\parallel^2 
+ \mu B + e\phi 
. 
\end{eqnarray}

Since the Lagrangian density 
${\cal L}_{DK} \equiv \int d^3 v \, F L_{GC}$ 
depends on time $t$ through not only 
the functions $(F, {\bf u}_x, u_\vartheta)$ determined by 
the variational principle but also
the given electromagnetic fields, 
$\delta_t I_{DK}$ does not vanish but it 
should be equal to 
\begin{equation}
\label{dLdt}
\epsilon \int_{t_1}^{t_2} \int_V d^3 x 
\int d^3 v \, F \left( \frac{\partial L_{GC}}{\partial t} \right)_u, 
\end{equation}
where $( \partial L_{GC}/\partial t )_u$ represents the derivative 
of  $L_{GC}$ in time $t$ with the variables 
$( {\bf u}_x, u_\vartheta )$ kept fixed. 
  Then, the local energy balance equation is derived 
from equating Eq.~(\ref{AdIDK2}) with Eq.~(\ref{dLdt}) and 
noting that the integral region $[t_1, t_2] \times V$ can be chosen arbitrarily. 
Besides, in the same way as explained in Sec.~V, 
we can see how the energy balance is modified 
when the term ${\cal K}$ representing collisions and/or 
external sources is added into the drift kinetic equation. 
  The resultant energy balance equation 
including the effect of ${\cal K}$ is written as 
\begin{eqnarray}
\label{energy_balance}
& & 
\frac{\partial }{\partial t} 
\left(
\int d^3 v \, F {\cal E} 
\right)
+ \frac{\partial }{\partial {\bf x} } 
\cdot 
\left(
\int d^3 v \, 
F {\cal E} {\bf u}_x
 \right)
\nonumber \\
& & 
=
\int d^3 v \, \left(  F   \dot{\cal E} 
+ {\cal K} {\cal E} 
\right) 
\end{eqnarray}
where the rate of change in the particle's energy is 
given by 
\begin{equation}
\dot{\cal E} 
\equiv 
- \left( \frac{\partial L_{GC}}{\partial t} \right)_u
= 
 e \frac{\partial \phi}{\partial t} 
+ \mu \frac{\partial B}{\partial t} 
- \frac{e}{c} 
{\bf u}_x  \cdot  \frac{\partial {\bf A}^* }{\partial t} 
.
\end{equation}
The energy balance equation shown in Eq.~(\ref{energy_balance}) 
agrees with Eq.~(11) in Ref.~\cite{Sugama2016}. 

We now consider the case of Sec.~IV where not only the distribution functions 
for all particle species but also the electromagnetic fields are determined 
by the governing equations which obey the variation principle. 
  Then, the variation in the action integral $I_{DKF}$ [see Eq.~(\ref{IDKF})] 
under the time translation is written as 
\begin{eqnarray}
\label{AdIDKF}
\delta_t I_{DKF} 
& = & 
\int_{t_1}^{t_2} \int_V d^3 x 
\left[ 
\epsilon \frac{\partial {\cal L}_{DKF} }{\partial t}
+ \sum_a  \int d^3 v 
\right. 
\nonumber \\ & & \mbox{}
\hspace*{-18mm}
 \times  \left\{
\delta_t F_a \cdot L_{GCa} 
+
 F_a \left(
\frac{\partial L_{GCa} }{\partial {\bf u}_{ax}} 
\cdot  \delta_t {\bf u}_{ax}
+ \frac{\partial L_{GCa}}{\partial  u_{a\vartheta}}  
 \delta_t u_{a\vartheta}
\right. \right. 
\nonumber \\ 
& & 
\hspace*{-18mm}
\left. \left. \left. 
\mbox{}
+ \frac{\partial L_{GCa} }{\partial \phi} \delta_t \phi
+ \frac{\partial L_{GCa} }{\partial {\bf A} } \cdot \delta_t {\bf A} 
\right) 
\right\}
-  \frac{\partial (B^2/8\pi) }{\partial (\partial {\bf A}/\partial x^i) } 
\cdot \frac{\partial (\delta_t {\bf A})}{\partial x^i}
\right]
, 
\nonumber \\ & & 
\end{eqnarray}
where $\delta_t {\bf u}_{ax}$, $\delta_t u_{a\vartheta}$, and 
$\delta_t F_a$ 
are given by using Eqs.~(\ref{Adu}) and (\ref{AdF}) for the particle 
species $a$ while $\delta_t \phi$ and $\delta_t {\bf A}$ are given by 
\begin{equation}
\delta_t \phi
=
- \epsilon \frac{\partial \phi}{\partial t}
,
\hspace*{5mm}
\delta_t {\bf A}
=
- \epsilon \frac{\partial {\bf A}}{\partial t}
.
\end{equation}
In the same way as in deriving Eq.~(\ref{AdIDK2}), 
we use the conditions for the particle's trajectory given by 
Eqs.~(\ref{dpidt}), (\ref{dLdv}), (\ref{dIdmu}), 
and (\ref{dIdth}) for each species $a$ as well as the 
additional conditions for the self-consistent fields 
given by $\delta I_{DKF} / \delta \phi =  0$ 
[Eq.~(\ref{quasineutrality})]
and $\delta I_{DKF} / \delta A_i =  0$
[Eq.~(\ref{ampere2})] in order to rewrite 
Eq.~(\ref{AdIDKF}) as 
\begin{equation}
\label{AdIDKF2}
\delta_t I_{DKF} 
=
- \epsilon 
\int_{t_1}^{t_2} \int_V d^3 x 
\left(
\frac{\partial {\cal E}_{tot} }{\partial t} 
+ \frac{\partial }{\partial {\bf x} } 
\cdot 
{\bf Q}_{tot}
\right)
, 
\end{equation}
where the total energy ${\cal E}_{tot}$ and the total energy flux 
${\bf Q}_{tot}$ are defined by 
\begin{equation}
\label{Etot}
{\cal E}_{tot} 
\equiv 
\sum_a \int d^3 v \, F_a 
\left( 
\frac{1}{2} m v_\parallel^2 
+ \mu B 
\right)
+ \frac{B^2}{8\pi}
, 
\end{equation}
and 
\begin{equation}
\label{Qtot}
{\bf Q}_{tot}
 \equiv 
\sum_a \int d^3 v \, F_a 
\left( 
\frac{1}{2} m v_\parallel^2 
+ \mu B  
\right) {\bf u}_{a x}
+ \frac{c}{4\pi}
( {\bf E} \times {\bf H} ) 
, 
\end{equation}
respectively. 
Here,  the magnetic intensity field ${\bf H}$ is defined by 
${\bf H} \equiv  {\bf B} - 4 \pi {\bf M}$ 
with the magnetic induction field ${\bf B}$ 
and the magnetization vector field 
${\bf M} \equiv (M^i)_{i=1,2,3}$ [see Eq.~(\ref{Mk})]
associated with the gyromotion of particles. 
It is noted that, 
in Eq.~(\ref{Etot}), 
the contribution of the electrostatic energy does not appear  
because $\sum_a \int d^3 v \, F_a e_a \phi = 0$ holds 
due to the quasineutrality condition. 
We also see from Eq.~(\ref{Qtot}) that 
the total energy flux ${\bf Q}_{tot}$ contains the kinetic energy flow 
due to the guiding center motion  
and the Poynting vector
$ (c / 4 \pi ) ( {\bf E} \times {\bf H} )$. 

Since the Lagrangian density ${\cal L}_{DKF}$ for the present system 
depends on time $t$ 
only through the distribution functions and the electromagnetic fields 
which are determined by the variational principle,  
the action integral $I_{DKF}$ given in Eq.~(\ref{IDKF}) is invariant 
under the time translation. 
   Therefore, noting that the integral region $[t_1, t_2] \times V$ 
can be arbitrarily chosen in Eq.~(\ref{AdIDKF2}), 
it follows that the integrand should vanish, 
which leads to the local energy conservation law. 
Furthermore, when the term ${\cal K}_a$ representing collisions 
and/or external sources is added into the drift kinetic equation for each 
particle species $a$, 
we can follow the procedure described in Sec.~V again to 
obtain the total energy balance equation, 
\begin{equation}
\label{total_energy_balance}
\frac{\partial {\cal E}_{tot} }{\partial t} 
+ \frac{\partial }{\partial {\bf x} } 
\cdot 
{\bf Q}_{tot}
=
\sum_a \int d^3 v \,  {\cal K}_a 
\left( 
\frac{1}{2} m v_\parallel^2 
+ \mu B 
\right)
, 
\end{equation}
where the condition 
$\sum_a e_a \int d^3 v \, {\cal K}_a = 0$ 
described after Eq.~(\ref{replace2}) in Sec.~V 
is used as well. 
The right-hans side of 
Eq.~(\ref{total_energy_balance}) vanishes when 
${\cal K}_a$ represents the collision operator 
which satisfies the conservation law of the kinetic energy.



\end{document}